\documentclass[a4paper,12pt,numbers,sort&compress]{article}

\pdfoutput=1

\usepackage[paper=letterpaper,margin=1in]{geometry}

\usepackage{amsmath,amssymb,amsfonts,epsfig,cite,setspace,bigstrut,longtable,array,breqn,pifont}

\renewcommand{\baselinestretch}{1.2}
\begin{document}


\vskip 0.25in

\newcommand{\todo}[1]{{\bf ?????!!!! #1 ?????!!!!}\marginpar{$\Longleftarrow$}}
\newcommand{\fref}[1]{Figure~\ref{#1}}
\newcommand{\tref}[1]{Table~\ref{#1}}
\newcommand{\sref}[1]{\S~\ref{#1}}
\newcommand{\nn}{\nonumber}
\newcommand{\tr}{\mathop{\rm Tr}}
\newcommand{\comment}[1]{}

\newcommand{\cA}{{\cal A}}
\newcommand{\cM}{{\cal M}}
\newcommand{\cW}{{\cal W}}
\newcommand{\cN}{{\cal N}}
\newcommand{\cH}{{\cal H}}
\newcommand{\cK}{{\cal K}}
\newcommand{\cZ}{{\cal Z}}
\newcommand{\cO}{{\cal O}}
\newcommand{\cB}{{\cal B}}
\newcommand{\cC}{{\cal C}}
\newcommand{\cD}{{\cal D}}
\newcommand{\cE}{{\cal E}}
\newcommand{\cF}{{\cal F}}
\newcommand{\cX}{{\cal X}}
\newcommand{\IA}{\mathbb{A}}
\newcommand{\IP}{\mathbb{P}}
\newcommand{\IQ}{\mathbb{Q}}
\newcommand{\IH}{\mathbb{H}}
\newcommand{\IR}{\mathbb{R}}
\newcommand{\IC}{\mathbb{C}}
\newcommand{\IE}{\mathbb{E}}
\newcommand{\IF}{\mathbb{F}}
\newcommand{\IV}{\mathbb{V}}
\newcommand{\II}{\mathbb{I}}
\newcommand{\IZ}{\mathbb{Z}}
\newcommand{\re}{{\rm Re}}
\newcommand{\im}{{\rm Im}}
\newcommand{\sym}{{\rm Sym}}

\newcommand{\tmat}[1]{{\tiny \left(\begin{matrix} #1 \end{matrix}\right)}}
\newcommand{\mat}[1]{\left(\begin{matrix} #1 \end{matrix}\right)}
\newcommand{\diff}[2]{\frac{\partial #1}{\partial #2}}
\newcommand{\gen}[1]{\langle #1 \rangle}
\newcommand{\ket}[1]{| #1 \rangle}
\newcommand{\jacobi}[2]{\left(\frac{#1}{#2}\right)}

\newcommand{\drawsquare}[2]{\hbox{%
\rule{#2pt}{#1pt}\hskip-#2pt
\rule{#1pt}{#2pt}\hskip-#1pt
\rule[#1pt]{#1pt}{#2pt}}\rule[#1pt]{#2pt}{#2pt}\hskip-#2pt
\rule{#2pt}{#1pt}}
\newcommand{\fund}{\raisebox{-.5pt}{\drawsquare{6.5}{0.4}}}
\newcommand{\antifund}{\overline{\fund}}

\newtheorem{theorem}{\bf THEOREM}
\def\thetheorem{\thesection.\arabic{theorem}}
\newtheorem{proposition}{\bf PROPOSITION}
\def\thetheorem{\thesection.\arabic{proposition}}
\newtheorem{observation}{\bf OBSERVATION}
\def\thetheorem{\thesection.\arabic{observation}}

\def\theequation{\thesection.\arabic{equation}}
\newcommand{\setall}{\setcounter{equation}{0}
        \setcounter{theorem}{0}}
\newcommand{\setequation}{\setcounter{equation}{0}}
\renewcommand{\thefootnote}{\fnsymbol{footnote}}

~\\
\vskip 1cm

\begin{center}
{\large \bf Calabi-Yau Geometries: Algorithms, Databases, and Physics}
\end{center}
\medskip

\vspace{.4cm}
\centerline{
{Yang-Hui He} \footnote{Invited review for the Int.~J.~of Modern Physics A and based on recent talks at CERN, Simons Center for Geometry \& Physics, Harvard University, McGill University, USTC China, the US Air Force Academy, and the Oxford James Martin School for the 21st Century.}
}
\vspace*{3.0ex}

\begin{center}
{\it
{\small
Department of Mathematics, City University, London, EC1V 0HB, UK; \\
School of Physics, NanKai University, Tianjin, 300071, P.R.~China; \\
Merton College, University of Oxford, OX14JD, UK\\
\qquad hey@maths.ox.ac.uk\\
}
}
\end{center}

\vspace*{4.0ex}
\centerline{\textbf{Abstract}} \bigskip
With a bird's-eye view, we survey the landscape of Calabi-Yau threefolds, compact and non-compact, smooth and singular.
Emphasis will be placed on the algorithms and databases which have been established over the years, and how they have been useful in the interaction between the physics and the mathematics, especially in string and gauge theories.
A skein which runs through this review will be algorithmic and computational algebraic geometry and how, implementing its principles on powerful computers and experimenting with the vast mathematical data, new physics can be learnt.
It is hoped that this inter-disciplinary glimpse will be of some use to the beginning student.

\newpage 
\renewcommand{\baselinestretch}{0}
\tableofcontents
\renewcommand{\baselinestretch}{1.2}


\section{Introduction}	
Whereas the archetype of the Renaissance scholar is that of a polymath versed in a multitude of subjects, and that of the Victorian inventor, a solitary figure labouring away in some focused esoterica, it is becoming ever clear that the scientist of the 21st century is once more obliged toward the former model.
With the increasingly blurring boundary between intellectual pursuits, the exponential growth of data and the rapidity of communication, inter-disciplinary research which harnesses the power of modern computing is assuming a steadily pre-dominant r\^ole.
Indeed, {\em CERN}, with its largest multi-national scientific collaboration, {\em PolyMath}, with its efficient utility of parallel minds blogging together \footnote{As an example, within a month of Zhang's seminal progress in attacking the Twin-Prime Conjecture, PolyMath has managed to lower the bound from $\sim 7\times 10^{7}$ to $\sim 10^4$. 
}
and {\em SVP}, with its global vision to systematically study the plethora of string vacua, all exemplify the forefront of this new paradigm.

Within the field of mathematical physics, especially in string theory, a success story particularly illustrative of this mode of theoretical research is that of Calabi-Yau spaces.
The story began in the late 1980's, when the high energy physics community was invigorated by the discovery of the ten-dimensional heterotic super-string, its natural incorporation of GUT-like gauge groups, and its potential to reach the low-energy, four-dimensional, Standard Model with particle generations upon compactification on Calabi-Yau threefolds \cite{Candelas:1985en}.
Thus arose a parallel challenge to physicists and mathematicians, in constructing such spaces and in translating the geometry into the physics.
Over the decades, the study of Calabi-Yau manifolds has blossomed into an incredibly rich subject, ranging from pure mathematics to particle phenomenology, allowing us to witness countless ground-breaking research in enumerative geometry, mirror symmetry, quiver representations, moduli spaces, dualities in QFT, a wealth of explicit gauge/gravity holographic duals, et cetera.

This success, and the numerous ones yet to come, certainly place Calabi-Yau manifolds as a central character upon the stage of modern theoretical science.
Confronted with the vastness of the subject, the limitations of space and knowledge clearly restrict me to a very specialized viewpoint of so breathtaking a landscape.
The perspective we will take is one from algorithmic and computational algebraic geometry.

This standpoint is compelled upon us physically and mathematically.
We now know there is an overwhelming degeneracy of possible string vacua, all of which resemble (but perhaps very few completely recover all the aspects of) the Standard Model.
Short of a {\it selection principle}, an immediate method of approach in isolating a particular compactification manifolds is not obvious. 
Instead, a synthetic rather than analytic perspective may prove to be conducive: could we attempt to establish large databases and develop efficient algorithms, and thereby ``experiment'', catalogue and analyse, in order to extract new physics and new mathematics?

As we shall see in the ensuing exposition, this philosophy was indeed the first course of action even in the infancy of the field, and has since matured fruitfully.
Of course, such a philosophy of algorithmic scanning and data mining is significantly facilitated by the rapid advances in computational algebraic geometry as well as its implementation on ever-faster machines, especially over the last decade \cite{sing,m2,GAP,sage,bertini}; the cross-pollination of large-scale computing and computational geometry with theoretical and mathematical physics has indeed recently been a healthy endeavour \cite{comp-book}.

Our review will proceed along this strand of thought and divides itself into three parts.
First, we survey the construction of smooth compact Calabi-Yau threefolds since the initial challenge three decades ago.
We will see how one can improve upon merely adhering to the tangent bundle by establishing more general stable bundles and how this leads to more salient phenomenology.
Second, we will investigate the space of non-compact, or local, Calabi-Yau threefolds which became a key player a decade after the 1980's when AdS/CFT brought the holographic principle and subsequently affine Calabi-Yau spaces to the limelight.
We conclude in the final part by turning to gauge theories in a context completely free of string theory, and will find surprising appearances of our familiar Calabi-Yau geometries.

\section{Triadophilia: CY3 and Stable Bundles}
\subsection{Prologue: a Three-Decade Search}
Our story begins with a thirty-year-old quest, which has prompted much activity over the decades and review some recent methodology and progress in addressing it.
The problem comes from string theory \cite{Candelas:1985en} , and constitutes the beginning of what we today call {\em string phenomenology}: the heterotic string gives a 10-dimensional supersymmetric theory with gauge group $E_8 \times E_8$, can one find a Calabi-Yau threefold, the compactification upon which will break one of the $E_8$ groups (the so-called ``visible'' \footnote{The other $E_8$ is called ``hidden'' and interacts via gravity mediation; we will not discuss the hidden sector here though much interesting phenomenology exist there as well \cite{Faraggi:2000pv,Braun:2006da,Braun:2013wr}.}) to something akin to the Standard Model group, together with particles and interactions familiar to our four-dimensions?

This has turned out to be a succinctly stated and well-motivated challenge to algebraic geometry.
The initial \cite{Greene:1986ar} realization was that the $SU(3)$ tangent bundle $TX$ of a Calabi-Yau threefold $X$ breaks the $E_8$ to the commutant $E_6$, whereby giving a four-dimensional supersymmetric $E_6$-GUT theory whose {\bf 27} representation, which endoes all the Standard Model fermions, is computed by the cohomology group $H^1(X,TX) \simeq H^{2,1}(X)$ and whose anti-generations of ${\bf \overline{27}}$ representations is computed by $H^1(X,TX^*) \simeq H^{1,1}(X)$; here we have used Hodge decomposition to relate the relevant groups to the familiar Hodge numbers.
Hence, that there should be three net generations of particles, is nicely summarized by the constraint
\begin{equation}\label{euler}
3 = \left| h^{1,1}(X) - h^{2,1}(X) \right| \quad \Rightarrow \quad
\chi(X) = \pm 6 \ ,
\end{equation}
where we have use the standard topological fact that for a Calabi-Yau threefold, the difference of the two Hodge numbers is half its Euler number $\chi(X)$.
Thus this so-called {\em triadophilia}, or the love of three-ness \cite{Candelas:2007ac}, phrased in terms of purely geometrical conditions, was born \footnote{
Several classical linguists, of which Oxford certainly has an abundance, were consulted, and we finally settled with the suggestion by Philip's daughter on this choice of the Greek.
Our love for ``three-ness'' is obvious, however, it would be a far greater desire to conceive of a geometrical {\it genesis} of this ``three-ness''.
}.

Of course, $E_6$ GUTs are less favoured today and our ultimate goal is to reach the (supersymmetric) Standard Model.
Group theoretically, this is a straight-forward three-step procedure \cite{Slansky:1981yr}:
\begin{enumerate}
\item We can use $SU(4)$ and $SU(5)$ to break the $E_8$ to the commutant $SO(10)$ and $SU(5)$ which are more popular GUTs (we include the $E_6$ case for reference)
\[
\mbox{
\begin{tabular}{|l|l|}
\hline
$E_{8}\rightarrow G\times H$ & Breaking Pattern \\  \hline\hline
$\rm{SU}(3)\times E_{6}$ & $248\rightarrow (1,78)\oplus (3,27)\oplus (\overline{
3}
,\overline{27})\oplus (8,1)$ \\  \hline
$\rm{SU}(4)\times\rm{SO}(10)$ &$ 248\rightarrow (1,45)\oplus (4,16)\oplus (\overline{4
},\overline{16})\oplus (6,10)\oplus (15,1)$ \\  \hline
$\rm{SU}(5)\times\rm{SU}(5)$ & $248\rightarrow (1,24)\oplus (5,\overline{10})
\oplus (
\overline{5},10)\oplus (10,5)\oplus (\overline{10},\overline{5})\oplus (24,1)$
 \\ \hline
\end{tabular}
}
\]
\item Next, we can use a Wilson line, which is a discrete finite group, typically $\IZ_k$ or $\IZ_k \times \IZ_{k'}$ to break the GUT to the Standard Model.
As canonical examples, for $SO(10)$ broken by a $\IZ_3 \times \IZ_3$ Wilson line to $SU(3) \times SU(2)_{U(1)_Y, U(1)_{B-L}}$, for the fermions and the Higgs, we have
\begin{align}
\nonumber
{\bf 16} &\rightarrow ({\bf 3},{\bf 2})_{(1,1)} \oplus ({\bf 1},{\bf 1})_{(6,3)}
\oplus ({\bf \overline{3}},{\bf 1})_{(-4,-1)} \oplus ({\bf \overline{3}},{\bf 1})_{(2,-1)} \oplus ({\bf 1},{\bf 2})_{(-3,-3)}\oplus ({\bf 1},{\bf 1})_{(0,3)}
\\
{\bf 10} &\rightarrow ({\bf 1},{\bf 2})_{(3,0)} \oplus ({\bf 3},{\bf 1})_{(-2,-2)}\oplus ({\bf 1},{\bf 2})_{(-3,0)} \oplus ({\bf 3},{\bf 1})_{(2,2)}
\end{align}
Similarly, for $SU(5)$ broken by an $\IZ_2$ Wilson line to $SU(3) \times SU(2)_{U(1)_Y}$, we have
\begin{align}
\nonumber
{\bf 5} &\rightarrow ({\bf 3},{\bf 1})_{-2} \oplus ({\bf 1},{\bf 2})_{3} \ , \quad
 {\bf \overline{5}} \rightarrow ({\bf \overline{3}},{\bf 1})_{2} \oplus ({\bf 1},{\bf 2})_{-3} \ ; \\
{\bf 10} &\rightarrow ({\bf 3},{\bf 1})_{4} \oplus ({\bf 1},{\bf 1})_{-6} \oplus ({\bf \overline{3}},{\bf 2})_{-1} \ , \quad
{\bf 10} \rightarrow ({\bf \overline{3}},{\bf 1})_{-4} \oplus ({\bf 1},{\bf 1})_{6} \oplus ({\bf 3},{\bf 2})_{1} \ .
\end{align}
Here, the Standard Model particles are (we include the extra $B-L$ charge), with requisite multiplicity (generation)
\begin{equation}\label{mssmfields}
\mbox{
\begin{tabular}{|l|l|l|}\hline
$({\bf 3,2})_{1,1}$ & $3$ & left-handed quark\\\hline

$({\bf 1,1})_{6,3}$ &$3$&left-handed anti-lepton\\\hline

$({\bf \overline{3},1})_{-4,-1}$&$3$&left-handed  anti-up\\ \hline

$({\bf \overline{3},1})_{2,-1}$&$3$&left-handed anti-down\\ \hline

$({\bf 1,2})_{-3,-3}$& $3$& left-handed lepton\\ \hline

$({\bf 1,1})_{0,3}$&$3$&left-handed anti-neutrino\\ \hline

$({\bf 1,2})_{3,0}$&$1$&up Higgs\\ \hline

$({\bf 1,2})_{-3,0}$&$1$& down Higgs\\\hline
\end{tabular}
}
\end{equation}
All other representations are {\it exotics}.

\item Finally, the Yukawa couplings are obtained by composing the appropriate triples of representations which give rise to gauge singlets.
\end{enumerate}

What is the geometrical structure which encapsulates the above group theory?
Over the years, collaborations betweens algebraic geometers and physicists have rephrased this as a clear problem \cite{Distler:1987ee,hubsch,Blumenhagen:1995ew,Friedman:1997ih,Friedman:1997yq,Donagi:1999ez,Donagi:2000zs,Donagi:2004ia,Donagi:2004ub,Donagi:2004su,Curio:2004pf,Andreas:2006zs,Blumenhagen:2005ga,Braun:2005zv,Weigand:2006yj,Distler:2007av}:
\begin{quote}
{\bf Challenge }
Does there exist a stable holomorphic vector bundle with structure group $G = SU(4)$ or $SU(5)$ on a smooth compact Calabi-Yau threefold $X$ with fundamental group $\Gamma = \pi_1(X)$ such that the relevant equivariant bundle cohomology group ($W$ is a representation of the $\Gamma$-Wilson line) $[ H^*(X, \bigwedge^q V^p \otimes W) ]^{\Gamma}$ carries the required particle representations above?
\end{quote}
More precisely, the cohomology groups $H^*(X, \bigwedge^q V^p)$, are
\begin{align}
\nonumber
G = SU(4) &: {\bf 16} = H^{1}(V) \ , \ {\bf \overline{16}} = H^{1}(V^*)\ , \ 
{\bf 10}= H^{1}(\wedge ^{2}V) \\
G = SU(5) &: {\bf 10} = H^{1}(V^*) \ , \ {\bf \overline{10}} = H^{1}(V)\ , \ 
{\bf 5}= H^{1}(\wedge ^{2}V) \ , {\bf \overline{5}}= H^{1}(\wedge ^{2}V^*) \ ,
\label{coho}
\end{align}
and the number of vector bundle moduli is given by $H^1(V\otimes V^*)$.
Indeed, the case of $V=TX$ returns us to the original case of \eqref{euler} and $E_6$ GUTs.

To answer this challege, we need three consecutive steps, which had been undertaken over the past 30 years, illustrating precisely the keywords of our title: new geometry, efficient algorithms and large databases:
\begin{enumerate}
\item Establish a ``landscape'' of smooth Calabi-Yau threefolds $X$;
\item Create databases of stable vector bundles $V$ on various families in $X$;
\item Develop techniques of computing cohomology group and trilinear (Yukawa) maps on a large scale.
\end{enumerate}
To each of these steps we shall focus, but first we have with this long physics prologue introduced our chief protagonist: Calabi-Yau threefolds. 
Therefore to these we will turn our present attention.

\subsection{Calabi-Yau Threefolds}
The definition of a Calabi-Yau manifold is by now familiar to a neophyte in theoretical physics: it is a (complex) K\"ahler manifold admitting flat Ricci curvature.
There are many equivalent definitions of which the above is the most intuitive.
Since the inclination of this review will be on the algebraic rather than the differential, perhaps the useful definition for us is that a Calabi-Yau manifold is
\begin{quote}
A complex algebraic variety with trivial canonical sheaf.
\end{quote}
Note that defined in such generality, we make no assumption whether the Calabi-Yau space is compact or not, singular or not.
Indeed, in the smooth compact case, the famous theorem of S.-T.~Yau states that the vanishing of the first Chern class guarantees the existence (and uniqueness) of such a flat K\"ahler metric.
We will also encounter non-compact and singular Calabi-Yau spaces; there, we understand the definition to be singularities which locally allow so-called {\em crepant resolutions} so that the resulting smoothed space has trivial canonical bundle.

The most famous Calabi-Yau threefold is indubitably the {\bf quintic} $Q$, so called because it is defined as a generic quintic polynomial in $\IP^4$.
This is a general lesson: a degree $d+2$ polynomial with sufficiently generic coefficients in $\IP^{d+1}$ (which has $d+2$ projective coordinates) will define a smooth Calabi-Yau $d$-fold.
That the sum of the degrees of the $d+2$ projective coordinates is equal to the degree of the defining polynomial implies the vanishing of the first Chern class.

\begin{figure}[t]
\begin{center}
(a)
\includegraphics[scale=0.35]{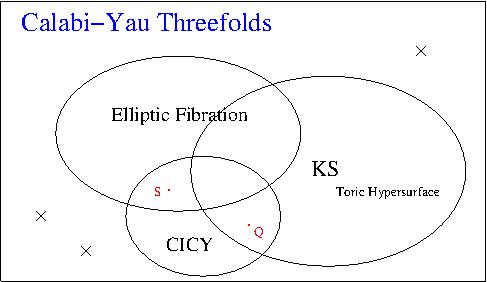}
(b)
\includegraphics[scale=0.45]{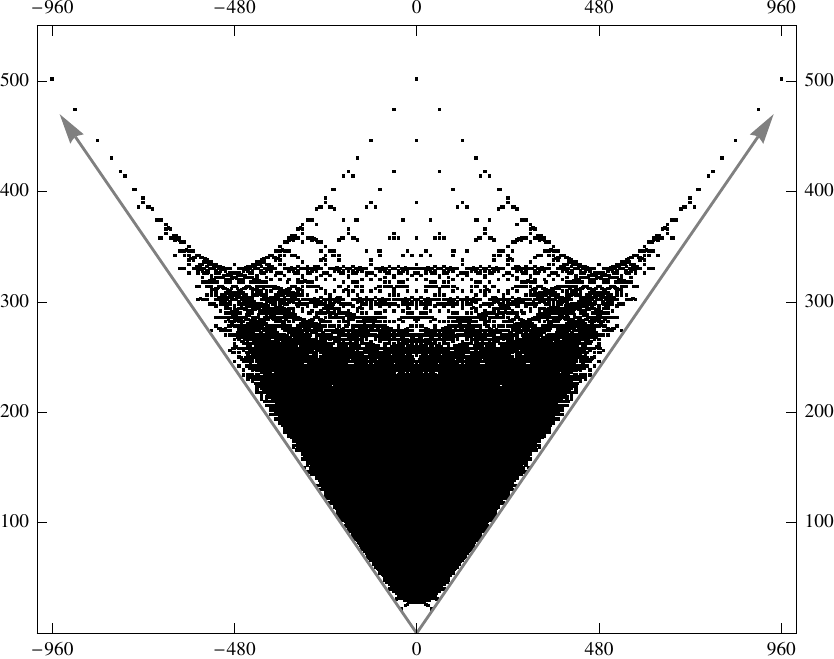}
(c)
\includegraphics[scale=0.25]{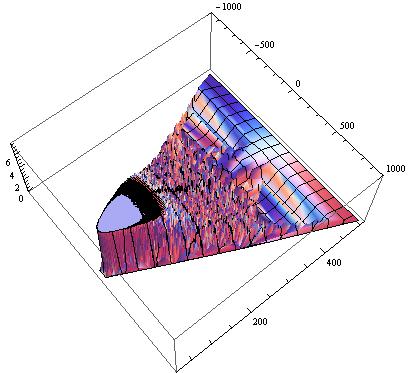}
\caption{{\sf 
(a) The space of CY3, with the 3 most studied datasets.
There are also some individualized constructions outside the three major databases, symbolically marked as crosses; these have been tabulated \cite{Candelas:2007ac,Candelas:2008wb,Qureshi:2011cg}.
$Q$ is the quintic, $S$ is the Schoen CY3 and the most ``typical'' CY3 has Hodge numbers $(27,27)$, totalling almost 1 million.
(b) Plotting $\chi = 2(h^{1,1}-h^{1,2})$ (horizontal) versus $h^{1,1} + h^{1,2}$ (vertical) of all the known Calabi-Yau threefolds.
(c) A refinement of (b) with the Log of frequency of the Hodge numbers in the vertical axis.
}}
\label{f:CY3}
\end{center}
\end{figure}

The construction of Calabi-Yau threefolds (CY3) has a distinguished history.
In Figure \ref{f:CY3} (a), we draw a (topogically correct but metrically non-representative) Venn diagram of some of the popular datasets thus far. In part (b) of the figure, we present the famous Hodge plot of $h^{1,1}(X) + h^{2,1}(X)$ in the ordinate versus $\chi = 2(h^{1,1}(X) - h^{2,1}(X))$ in the abscissa; the apparent left-right symmetry of the diagram is the best experimental evidence for {\it mirror symmetry}.
In part (c), we also indicate the Log of the frequency of the Hodge numbers: indeed there is tremendous redundancy, the some $10^9$ CY3 (we will see in \S\ref{s:KS} that there is much more than $10^9$ known CY3s though their full characterization still awaits work) share only about $10^5$ Hodge pairs and interestingly the most pre-dominant pair known so far is $(h^{1,1}, h^{2,1}) = (27,27)$, totalling 910113.

Amusingly, the largest known magnitude of Euler number of any CY3 is 960, corresponding to the mirror Hodge pairs $(11,491)$ and $(491,11)$.
This is also twice the difference between the dimension of the adjoint and the rank of $E_8 \times E_8$.
Incidentally, for the reader's further digression, twice the dimension, $248 \cdot 2 = 496$ is the only perfect number in the hundreds.

\subsubsection{CICY Manifolds}
To address \eqref{euler}, the first database of CY3 was the so-called {\bf CICY manifolds}\cite{cicy,hubsch}, or complete intersection Calabi-Yau threefolds embedded as $K$ homogeneous polynomials in $\IP^{n_1} \times \ldots \times \IP^{n_m}$. 
This is clearly a direct generalization of the quintic.
Here, complete intersection means that the dimension of the ambient space exceeds the number $K$ of defining equations by precisely 3, i.e., $K=\sum\limits_{r=1}^m n_r-3$. Moreover, the Calabi-Yau condition of vanishing first Chern class of $TX$ translates to $\sum\limits_{j=1}^K q^{r}_{j} = n_r + 1 \ \forall \; r=1, \ldots, m$.
Subsequently, each manifold can be written as an $m \times K$ configuration matrix (to which we may sometimes adjoin the first column, designating the ambient product of projective spaces, for clarity):
\begin{equation}\label{cicy}
X = 
\left[\begin{array}{c|cccc}
  \IP^{n_1} & q_{1}^{1} & q_{2}^{1} & \ldots & q_{K}^{1} \\
  \IP^{n_2} & q_{1}^{2} & q_{2}^{2} & \ldots & q_{K}^{2} \\
  \vdots & \vdots & \vdots & \ddots & \vdots \\
  \IP^{n_m} & q_{1}^{m} & q_{2}^{m} & \ldots & q_{K}^{m} \\
  \end{array}\right]_{m \times K \ ,}
\quad
\begin{array}{l}
K = \sum\limits_{r=1}^m n_r-3 \ , \\
\sum\limits_{j=1}^K q^{r}_{j} = n_r + 1 \ , \ \forall \; r=1, \ldots, m \ .
\end{array}
\end{equation}
The Chern classes of $X$ are (of course, $c_1^r(TX) = 0$):
\begin{equation}
c_2^{rs}(TX) = \frac12 \left[ -\delta^{rs}(n_r + 1) + 
  \sum_{j=1}^K q^r_j q^s_j \right] \ , \quad
c_3^{rst}(TX) = \frac13 \left[\delta^{rst}(n_r + 1) - 
  \sum_{j=1}^K q^r_j q^s_j q^t_j \right] \ , 
\label{CICY-chernX}
\end{equation}
where we have written the coefficients of the total Chern class
$c = c_1^r J_r + c_2^{rs} J_r J_s + c_3^{rst} J_r J_s J_t$ explicitly, with $J_r$ being the K\"ahler form in $\IP^{n_r}$.
The triple-intersection form $d_{rst} = \int_X J_r \wedge J_s \wedge J_t$ is a totally symmetric tensor on $X$ and the Euler number is simply $\chi(X) = d_{rst} c_3^{rst}$.

The construction of CICYs was thus reduced to a combinatorial problem of classifying the integer matrices in \eqref{cicy}.
It was shown that such configurations were finite in number and the best available computer at the time (1990's) was employed, viz., the super-computer at CERN \cite{cicy,hubsch}.
A total of 7890 manifolds were found, including, of course, our quintic in $\IP^4$, which we can now write as $Q =  [4|5]_{-200}^{1,101}$, where we have written the Hodge numbers and Euler number respectively as super- and sub-scripts.
Another famous CICY is the Schoen manifold, $S = {\small \mat{1&1\\3&0\\0&3}_0^{19,19}}$, which is a self-mirror CY3.
We mark these two red points in (a) of Figure \ref{f:CY3} and will return to address them.
Of these some 8000 threefolds, unfortunately none has $\chi = \pm 6$, which was an initial disappointment. Of course, today, our generalization from $TX$ to $V$ no longer has \eqref{euler} as a triadophilic constraint.
Nonetheless, the transpose of $S$ was soon found by Tian and Yau to have a freely acting $\IZ_3$ symmetry, so that the quotient $M = \mat{1&3&0\\ 1&0&3\\} / \IZ_3$  has topological numbers $M_{-6}^{6,9}$ which did cause a sensation at the time.

A few points are worthy of note.
The transpose configuration of a CICY is also a CICY and constitutes, in fact, a conifold transition \cite{hubsch,Candelas:2007ac}.
When a CY3 has $h^{1,1} = 1$, it is called {\em cyclic}.
There are only 5 cyclic CICYs, viz.,
$Q=[4|5], [5|3,3], [5|2,4], [6|2,2,3], [7|2,2,2,2]$. 
The transposes of these are thus also CICY and we will denote them as cyclic$^T$.

\subsubsection{Elliptic CY3}\label{s:elliptic}
As the CICY manifolds dominated the late 1980's for a number of years, from the interest in F-theory in the late 1990's emerged another dataset\cite{Morrison:1996na,Grassi:2000we,Friedman:1997yq,Donagi:1997dp} of CY3, viz., those which are elliptically fibred over some complex surface $B$.
Over $B$ the CY3 is realized as a possibly degenerate torus with section $\sigma$ and can thus be realized as an elliptic curve.

The existence of the section highly constrains what $B$ could be \cite{Morrison:1996na}, being only one of the following:
\begin{enumerate}
\item Hirzebruch surfaces $\IF_r$ for $r = 0, 1, \ldots, 12$;
\item $\IP^1$-blowups of Hirzebruch surfaces $\widehat{\IF}_r$ for $r=0,1,2,3$;
\item Del Pezzo surfaces $d\IP_r$ for $r = 0,1,\ldots,9$;
\item Enriques surface $\IE$.
\end{enumerate}
The $F_r$ are various ways which $\IP^1$ could fibre over $\IP^1$.
The Enriques surface is a $\IZ_2$ quotient of K3 and to the del Pezzo surfaces we will return in \S\ref{s:dP}.

In terms of the Chern classes of the tangent bundle $TB$ of the base, we have
(of course, $c_1(TX) = 0$):
\begin{equation}\label{elliptic-chernX}
c_2(TX) = \pi^*(c_2(TB) + 11 c_1(TB)^2) + 12 \sigma \cdot \pi^*(c_1(TB)) \ ,
\quad
c_3(TX) = -60 c_1(TB)^2 \ ,
\end{equation}
where $\pi : X \rightarrow B$ is the projection map of the elliptic fibration.
Even though the list of possible bases seems limited, by tuning the possible elliptic curve, an incredibly diverse range of CY3 can be reached.
Of the known CY3, many tens of thousands have been identified as elliptic fibrations \cite{Braun:2011ux,Taylor:2012dr}; the full classification of this rich dataset is still in progress.

\subsubsection{Kreuzer-Skarke List}\label{s:KS}
The largest set of CY3 known today is due to many years of impressive work by Kreuzer-Skarke (KS); these are the hypersurfaces in toric varieties of dimension four\cite{BB,Avram:1997rs,Kreuzer:2000qv}.
This is an extensive generalization of the CICYs with $K=1$ by having, as ambient space, not merely products of projective spaces (in dimension 4 there are only 5, corresponding to the 5 partitions of 4, $\IP^4$, $\IP^2 \times \IP^2$, $\IP^3 \times\IP^1$, $\IP^1 \times \IP^1 \times \IP^2$, and $(\IP^1)^4$; these give, of course, precisely our cyclic$^T$ CICYs).
Indeed, one can think of hypersurfaces\cite{Candelas:1989hd} in weighted $\IP^4$, of which there are about 6000, as a nice intermediate step.

The general construction is elegant and combinatorial.
Take a polytope $\Delta \in \IR^4$ with integer vertices which contains the origin and consider its dual $\Delta^\circ = \{\vec{v} \in \IR^4 : \vec{m} \cdot \vec{v} \ge -1, \forall \vec{m} \in \Delta \}$.
Each defines a toric variety in a standard way.
Now, $\Delta$ is called reflexive if $\Delta^\circ$ also has integer vertices, in such a case, the (shifted) Newton polynomial of $\Delta$, defined as $P(\vec{x}) = \sum\limits_{\vec{m} \in \Delta} C_{\vec{m}} \prod\limits_{i=1}^4 {x_i}^{\vec{m} \cdot \vec{v}_i +1}$ where $\vec{v}$ are the integer vertices of $\Delta^\circ$ and $C_{\vec{m}}$ are generic complex coefficients, defines a CY3 hypersurface.
Therefore, the classification KS CY3s amounts to that of reflexive integer 4-polytopes; in dimensions one to three these total 1, 16, 4319, respectively and the present case of dimension 4 was a major computational challenge.
The actual calculation was performed on an SGI origin 2000 machine with about 30 processors and took approximately 6 months and 473,800,776 was found.

\begin{figure}[t]
\begin{center}
(a)
\includegraphics[scale=0.5]{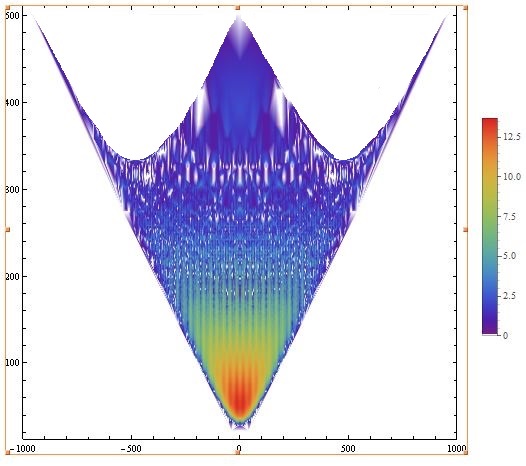}
(b)
\includegraphics[scale=0.3]{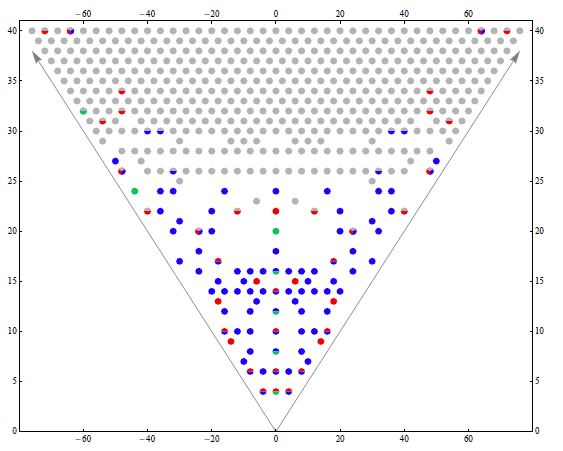}
\caption{{\sf 
(a) A colour Log-density plot of the Hodge numbers of the KS dataset, with  $\chi = 2(h^{1,1}-h^{1,2})$ (horizontal) versus $h^{1,1} + h^{1,2}$ (vertical).
(b) A zoom-in of the tip with small Hodge numbers: the gray are KS CY3s and the coloured are individually engineered cases; this plot is taken from Candelas-Davies\cite{Candelas:2008wb}.
}}
\label{f:KSplots}
\end{center}
\end{figure}

We need to emphasize a few points.
These polytopes correspond to possibly singular 4-folds (in fact, only 124 are smooth\cite{He:2009wi}), thus the majority thereof requires desingularization by triangulation, standard to toric geometry.
To each desingularization we can associate a new hypersurface and therefore the number of CY3 far exceeds $\sim 5 \times 10^9$.
Nevertheless, the Hodge numbers are invariants under the triangulations (the intersection form, however, would be different) and 30,108 distinct Hodge pairs have been found.
We show the Log-density plot of these Hodge numbers in Part (a) of Figure \ref{f:KSplots}.

Of course, the KS dataset, as large as it is, is only the tip of an iceberg, one could go on to study complete intersections in higher dimensional toric varieties, much like the CICY case.
The situation of the double-hypersurface in toric 5-folds was already nearing completion circa Max Kreuzer's untimely death \footnote{
I have a profound respect for Max.
It was not long after his visit to Oxford, a very productive and convivial period, that we received the shocking email that his doctors said he only had a few months left. During the last weeks on his deathbed as cancer rapidly took hold of him, he emailed us regularly and our many discussions continued as normal. The several posthumous papers on the ArXiv are testimonies to his dedication during his last hours; only a true scholar could have the courage of such extraordinary devotion. {\it In pace requiescat!}
}
in 2010.

There are beautiful patterns in the distribution of the Hodge numbers which still elude us today and many intriguing properties have been uncovered \cite{Cicoli:2011it,Taylor:2012dr,Candelas:2012uu}.
A particularly salient feature is that the ``tip'' of the plot is almost empty (considering the millions in the centre); the paucity of CY3 with small Hodge numbers, which also include all the manifolds which have become of phenomenological interest, suggests a possible oasis in the landscape of compactifications \cite{Candelas:2007ac,Candelas:2008wb,Braun:2009qy,He:2010uj}.

\subsection{Stable Vector Bundles}
Having taken a brief stroll within the realm of compact CY3 manifolds, we now need to move onto the next part of our survey, the bundle $V$ over $X$.
Some immediate constraints can be imposed:
\begin{itemize}
\item {\em Supersymmetry and stability: }
Requiring $\cN = 1$ SUSY in the low-energy 4-dimensional theory implies that $V$ admits a holomorphic connection $F$ satisfying the Hermitian-Yang-Mills (HYM) equations: $ F_{ab} = F_{\bar{a}\bar{b}} = g^{a\bar{b}}F_{a\bar{b}} = 0$, a generalization of Ricci-flatness for Calabi-Yau manifolds.
These impossibly difficult non-linear PDEs can be circumvented by the celebrated Donaldson-Uhlenbeck-Yau theorem (DUY) which states that on each (poly-)stable holomorphic vector bundle, there exits a unique HYM connection.

We focus on special unitary bundles here so $c_1(V) = 0$ and stability is the {\it algebraic} condition that there exists a K\"ahler class $J$ such that for every subsheaf $\cF$ of $V$, the quantity (called slope) $\mu(\cF) = \int_X c_1(\cF)J^2 < 0$.
The difficulty, therefore has been shifted to finding all subsheafs, which in many cases again becomes a problem in combinatorial algorithmics.
One nice consequence of stability for $SU(n)$ bundles is that, together with Serre duality on $X$,
\begin{equation}\label{stab}
H^0(X,\wedge^i V) = H^0(X, \wedge^i V^*) = H^3(X, \wedge^i V) = H^3(X, \wedge^i V^*) = 0 \ ,
\end{equation}
for all $i = 0, \ldots, n-1$.

\item {\em Triadophilia and Index Theorem: }
The vanishing conditions \eqref{stab}, together with the Atiyah-Singer index theorem on $X$ imply that
\begin{equation}
\mbox{index}(\slash \!\!\!\!\nabla_X) = \sum\limits_{i=0}^3 (-1)^i
h^i(X,V) = \int_X \text{ch}(V) \text{td}(X) = \frac12 \int_X c_3(V) \ .
\end{equation}
Consequently, this gives us an expression for the net number of generations of particles, generalizing \eqref{euler}:
\begin{equation}
N_{gens} = -h^1(X,V) + h^1(X,V^*) = \frac12 \int_X c_3(V) = 3k\ , \qquad
\chi(X)\bmod k = 0 \ ,
\end{equation}
where $k \in \IZ_{>0}$ is the order of a possible freely acting group $G$ on $X$, so that upon descending to the quotient manifold $X/G$, there would be precisely 3 generations.
Indeed, in order that $G$ be a free action, $k$ must necessarily (but not sufficiently) divide the Euler number $\chi(X)$.

\item {\em Anomaly Cancellation: }
To ensure Green-Schwarz anomaly cancellation, it is standard to set $\int_X R \wedge R - F \wedge F = 0$, where $R$ is the Ricci form on $X$, that is, $c_2(X) = c_2(V)$.
However, one could allow M5-branes in the bulk, in a heterotic M-theory Ho\v{r}ava-Witten set-up \cite{Horava:1996ma,Donagi:1999ez}, which could wrap effective holomorphic 2-cycles (i.e., actual, physical, curves).
Hence, anomaly cancellation requires that
\begin{equation}
c_2(X) - c_2(V) = \mbox{ effective class in } H_2(X;\IZ) \ .
\end{equation}
\end{itemize}

After imposing these preliminary conditions, we can then proceed to computation of the cohomology groups in \eqref{coho}; before doing so, it is expedient to follow the similar vein above and peruse over the available datasets.

\subsubsection{CICY Monads}
Since CICYs (in particular the quintic) provided the first database, the immediate next step was to construct bundles thereon.
Historically, this was indeed the case \cite{Distler:1987ee} and recently a programme was resurrected to systematically study such bundles \cite{Anderson:2007nc,Anderson:2008uw,Anderson:2009ge,Anderson:2009mh,Anderson:2012yf,Anderson:2011ns,Anderson:2013xx,Anderson:2009nt} using advanced computing and novel algebro-geometric algorithms \cite{m2,sing}.

The most appealing property of a CICY is its explicit projective coordinate, and thence, the description of line-bundles.
In general, our ambient space $\cA$ for CICYs is a product of $m$ projective 
spaces in which $K$ homogeneous polynomials define $X$. 
We shall call the situation where $h^{1,1}(X) = h^{1,1}(\cA) = m$ as {\em favourable}; here the K\"ahler classes descend completely from $\cA$ to $X$.
In this case we can write line bundles over $\cA = \mathbb{P}^{n_1 }\times \mathbb{P}^{n_2} \times \ldots \times \mathbb{P}^{n_m}$ as $\cO_{\cA}(k_1,k_2,...,k_m
)$ with corresponding restriction to $X$.
Equipped with line bundles, a natural (and indeed historical) next step is to construct so-called monads.

In general, a monad bundle \cite{monad} is the cohomology of the (non-exact) sequence $0 \to A \to B \to C \to 0$, with $A,B,C$ direct sums of line bundles;
for simplicity we take $A$ to be trivial and our monad bundle $V$ to reside in the short exact sequence
\begin{equation}
0 \to V \stackrel{f}{\longrightarrow} B \stackrel{g}{\longrightarrow}C \to 0 \ ;
\mbox{ with }
\qquad
{\scriptsize
B=\bigoplus\limits_{i=1}^{r_B} \cO(b_{r}^{i})\; ,\qquad
C=\bigoplus\limits_{j=1}^{r_C} \cO(c_{r}^j) \ .
}
\end{equation}
Here, short exactness implies that $V = \im(f) \simeq \ker(g)$ and that $\mbox{rk}(V)=\mbox{rk}(B)-\mbox{rk}(C)$.
The map $g$ is explicitly a matrix of polynomials; e.g., on $\mathbb{P}^n$ the $ij$-th entry is a homogeneous polynomial of degree $c_i - b_j$.

Our above physical constraints readily manifest themselves as a list of combinatorial conditions on the integers $b^i_r, c^j_r$ ($d^{rst}$ are the triple intersection numbers):
\begin{enumerate}
\item Bundle-ness: $b^{i}_r \leq c^{j}_r$ for all $i,j$ and the map $g$ can be taken to be {\it generic} so long as exactness of the sequence is ensured;
\item SU-Bundle: $c_1(V)=0  \Leftrightarrow \sum \limits_{i=1}^{r_B} b^{r}_i -\sum \limits_{j=1}^{r_C} c^{r}_j = 0$;
\item Anomaly cancellation: $c_2(X)- c_2(V) = c_2(X)-\frac12 (\sum\limits_{i=1}^{r_B} {b^{i}_s}{b^{i}_t} - \sum\limits_{i=1}^{r_C} {c^{j}_s}c^{t}_i) d^{rst} \geq 0$;
\item Three Generations: $c_3(V) =\frac{1}{3} (\sum\limits_{i=1}^{r_B}{{b_r}^i}{{b_s}^i}{{b_t}^i} -  \sum\limits_{j=1}^{r_C}{{c_r}^j}{{c_s}^j}{{c_t}^j}) d^{rst} = 3k$ \ .
\end{enumerate}
Once more, we witness a natural course of action: physics to algebraic geometry to combinatorics to computerization.

We remark that, much like the classification of CICYs, if we impose that all entries of $B$ and $C$ be positive, then one can show that the space of such monads on (favourable) CICYs is finite (some 7 thousand).
Of course, having non-positive entries is perfectly allowed and could lead to good models.
For example, recently, the $SU(4)$ monad bundle
\begin{equation}
0 \to V \to \cO_X(1,0)^{\oplus 3} \oplus \cO_X(0,1)^{\oplus 3} \stackrel{f}{\rightarrow} \cO_X(1,1) \oplus \cO_X(2,2) \to 0 \ ,
\end{equation}
defined on the bi-cubic CY3, $X=$  {\tiny $\left[\begin{array}[c]{c}\mathbb{P}^2\\\mathbb{P}^2\end{array}\left|\begin{array}[c]{ccc}3 \\3 \end{array}\right.  \right]^{2,83}_{-162}$} (which is a conifold transition of the Schoen and Tian-Yau manifolds) has been found to give the {\it exact} spectrum of the MSSM upon a $\IZ_3^2$ Wilson line \cite{Anderson:2009mh,Anderson:2009ge}.

\subsubsection{Spectral Covers}
The difficult part of the monad bundles is to prove stability; this has not yet been been fully automated.
To circumvent this, we turn to the elliptic database, for which a so-called {\em spectral cover} construction guarantees (for sufficiently large fibre class as polarization) stability \cite{Donagi:1997dp,Friedman:1997ih,Friedman:1997yq}.
Thereby one can obtain the largest explicit set of stable $SU(n)$ bundles \cite{Donagi:2004ia,Gabella:2008id} (about $10^7$).
Such a bundle is given by the spectral data, consisting of the following two pieces~:
\begin{itemize}
\item The {\em spectral cover} $\mathcal{C}_V$~:  this is an $n$-fold cover of the base and is thus a divisor (surface) in $X$ with degree $n$ over $B$, as an element in $H_4(X; \IZ) \simeq H^2(X,\mathbb{Z})$ it is $[\mathcal{C}_V] =  n\ \sigma +  \pi^* \eta$, where $\sigma$ is the class of the zero section, and $\eta$ is an effective curve class in $H^2(B,\mathbb{Z})$.
In order that $V$ be stable, $\cC$ needs to be irreducible, which follows from the constraints that
(a) the linear system $|\eta|$ is base-point free in $B$ and
(b) $\eta - n c_1(B)$ is an effective curve in $B$.

\item The {\em spectral line bundle} $\mathcal{N}_V$~: this is a line bundle on $\mathcal{C}_V$ with first Chern class
$c_{1}(\cN_V)=n(\frac{1}{2}+\lambda)\sigma+(\frac{1}{2}-\lambda) \pi^{*}\eta+(\frac{1}{2}+n\lambda)\pi^{*}c_{1}(B)$.
The parameter $\lambda$ has to be either integer or half-integer depending on the rank $n$ of the $SU(n)$ structure group~:
\begin{equation}
\lambda =  \left\{
\begin{array}{cl}
        m+1/2 & \text{ if $n$ is odd}, \\
        m & \text{ if $n$ is even},
\end{array} \right.
\end{equation}
where $m\in\mathbb{Z}$.
When $n$ is even, we must also impose $\eta = c_1(B) \textrm{ mod } 2$, 
by which we mean that $\eta$ and $c_1(B)$ differ only by an even element of $H^2(B,\mathbb{Z})$.
\end{itemize}

The Mori cone of effective curves on $X$ is spanned by 
\begin{equation}
\sigma \cdot \pi^*(C_i) \ , F \in H_2(X; \IZ) \ ,
\end{equation}
where $F$ is the fibre class and $C_i$ are a basis of effective curves in the base $B$.
The relevant intersections are (curve with surface and surface with surface): 
\begin{equation}
\sigma \cdot F = 1 \ , \quad
\pi^*(C_i) \cdot F = 0 \ , \quad
\pi^*(C_i) \cdot \pi^*(C_j) = \pi^*(C_i \cdot C_j) = C_i\cdot C_j F \ , \quad
\sigma \cdot \sigma = \sigma \pi^*(- c_1(TB)) \ , 
\end{equation}
where for the second intersection one uses the fact that one can always choose a fibre which generically misses a pull-back of a curve in the base;
for the third one uses intersection $C_i \cdot C_j$ in the base, giving a point, which then pulls back to a generic fibre; for the last, one uses adjunction.

The holomorphic $SU(n)$ vector bundle $V$ on $X$ can be extracted from the above data by a  Fourier-Mukai transformation: $(\mathcal{C}_V,\mathcal{N}_V) \,\stackrel{FM}{\longleftrightarrow}\, V$.
The Chern classes of $V$ are given in terms of the spectral data as (again, $c_1(V)=0$)
\begin{eqnarray}
\nn
   c_2(V) &=& \sigma \cdot\pi^*\eta 
              + \pi^*\left( 
              \frac{n}{2} \left(\lambda^2 - \frac 1 4\right) \eta \cdot
                 \left(\eta - nc_1(B)\right) - \frac{n^3 - n}{24} c_1(B)^2
              \right) := \sigma \pi^*\eta + c_F F   \ ,  \\
\label{cVspec}
   c_3(V) &=& 2 \lambda \eta \cdot \left(\eta - nc_1(B) \right) \ .  
\end{eqnarray}

A combination of spectral cover techniques and simple extensions of bundles over elliptic CY3 has met reassuring success and produced the first known answers to the challenge laid out in the beginning of this section.
On two different quotients of the aforementioned Schoen manifold $S$, stable bundles were constructed so as to give the {\it exact} particle content of the MSSM, together with reasonable Yukawa couplings \cite{Braun:2005nv,Bouchard:2005ag}.

\subsubsection{Polystable Bundles}
One of the most fruitful and perhaps also the easiest set of bundles is simply the direct sum of line bundles.
This is slightly different from the $SU(n)$ bundles we have so far been describing and are $S(U(1)^n)$ bundles which can be seen as the ``splitting'' of the former along walls of marginal stability within moduli space \cite{Anderson:2012yf,Anderson:2011ns,Anderson:2009nt,He:2012iy,Anderson:2013xx}. 
That is, our bundles are of the form $V = \bigoplus\limits_{i=1}^5 L_i$.

In the context of DUY, these are polystable bundles and can indeed also admit Hermitian Yang-Mills connection.
The breaking pattern is a little more complicated, for example, an $S(U(1)^5$ bundle will break $E_8$ down to an $SU(5) \times S(U(1)^5$ GUT.
A Wilson line can then break the $SU(5)$ into a Standard Model group, together with the extra massive Abelian factor $S(U(1)^5$.
Thus, though not minimal in having these extra $U(1)$ factors, one could still use traditional Green-Schwarz mechanism to let these acquire D-terms with FI parametres and whence masses.

Such (equivariant) polystable bundles having, up to these $U(1)$ factors, xact MSSM spectrum and reasonable Yukawa couplings, have been classified over the CICY database and very nicely give rise to the largest known set of heterotic Standard Models (some $10^5$) \cite{Anderson:2011ns,Anderson:2012yf,Anderson:2013xx}.
Obviously, the components of the line bundles $L_i$ must have mixed positive and non-positive entries in order to admit solutions of polarizations $J$ such that $\mu(V) = \mu(L_i) = 0$. 
In other words, polystability translates to the simple algebraic system $t^r \in \IR_{>0} : d_{rst} (L_i)_r t^s t^t = 0$ for all $i=1,\ldots,n$ where $d_{rst}$ are the triple intersection numbers, $(L_i)_r$ are the entries to the line bundles and $t^r$ are the K\"ahler parametres.

For example \cite{Anderson:2011ns}, on the tetraquadric CY3 (one of the aforementioned 5 that are both KS and CICY) $X = {\scriptsize \left[ 
\begin{array}{c|c}
\IP^1 & 2 \\\IP^1 & 2 \\\IP^1 & 2 \\\IP^1 & 2 \\
\end{array}
\right]_{-128}^{4,68}}$, $V = \cO_X(1,-3,0,2) \oplus \cO_X(0,1,0,-1) \oplus \cO_X(0,1,0,-1)\oplus \cO_X(0,0,-1,1)\oplus \cO_X(-1,1,1,-1)$ is an $\IZ_2^2$-equivariant $S(U(1)^5)$ bundles which gives, up to the $U(1)$ groups, exact MSSM spectrum.

\subsubsection{KS Bundles}
Of course, the largest gold mine of CY3 still awaits us.
To study the distribution and frequency of exact, and not merely quasi, Standard
 Models by systematically constructing stable bundles on at least half-billion manifolds 
\footnote{As discussed above, due to triangulations the actually number is far more than half billion and the full catalogue is in progress.} 
and computing their cohomology is under way.
This is clearly a task for large-scale parallel computing.

There have been some preparatory works toward this vision.
The positive monads on all KS CY3 which have smooth ambient space have been completely classified \cite{He:2009wi} and as a test-run, marching upward in $h^{1,1}$, the small values have also been addressed \cite{He:2011rs}.
In parallel, the requisite geometrical data (beyond the topological quantities such as Hodge numbers and Chern classes) such as intersection form and Mori cone of the full KS list are currently being computed \cite{Gray:2012jy,Gao:2013pra}.

Furthermore, anticipating the cohomology computation, a very nice computer package for calculating cohomology of line bundles on toric varieties has been written \cite{Blumenhagen:2011xn}.
Finally, the Sage project, which is an increasingly popular attempt to interface the multitude of computer software for mathematics, is becoming ever prominent and useful in our and much more general calculations \cite{sage}.

\section{Gauge Theory: CY3 and Quivers}
We have taken a rapid survey of the space of known compact CY3 and a glimpse of the large number of stable bundles for the sake of heterotic compactification, a challenge posed in the mid-1980's.
Now let us move on to a complementary view that arose in the mid-90's: instead of having 6 tiny extra dimensions, can we live on the world-volume of a brane floating transverse to 6 large extra dimensions?
This of course is the brane-world scenario, a brain-child of Maldecena's seminal work\cite{Maldacena:1997re} on AdS/CFT, highlighting a {\it holographic} principle which by now has extended far beyond string theory and rests as a corner-stone of 21st century physics.

In this set-up, there is a natural bijection between (1) the world-volume physics, which, for a stack of D3-branes, is some super-conformal four-dimensional gauge theory (SCFT) and (2) the transverse or ``bulk'' geometry, which is generically some non-compact CY3.
The asymptotic metric for the brane is anti-de-Sitter while the CY3 is a real cone over some 5-dimensional Sasaki-Einstein (SE) manifold.
By construction, the low-energy vacuum moduli space (VMS) parametrized by the scalars in the SUSY multiplets of our gauge theory is the CY3 since the motion of the branes is realized by the CY3 and parametrizes the geometrical degrees of freedom for the VMS.
Thus proceeding from (1) to (2) is the calculation of the F- and D-flatness conditions of a SUSY gauge theory (to this point we shall return in the last section) which gives the CY3 as a vacuum manifold and from (2) to (1), has been dubbed ``geometrical engineering'' \cite{Katz:1996fh}.

\begin{figure}[t]
\begin{center}
(a)
$\begin{array}{l}\includegraphics[scale=0.3]{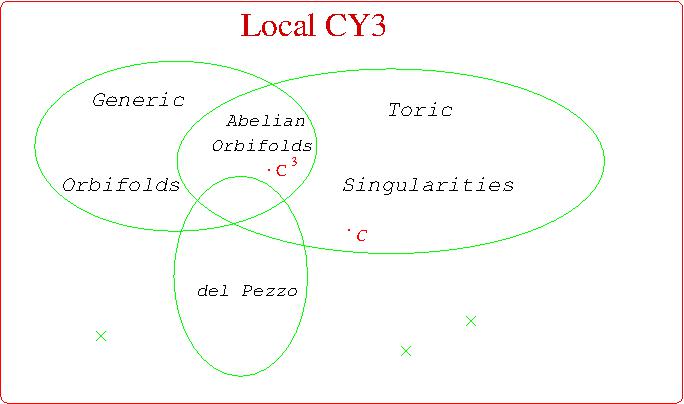}\end{array}$
(b)
$
\begin{array}{c}
\begin{array}{c}\includegraphics[scale=0.6]{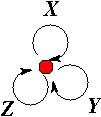}\end{array} \\
W = \tr(X[Y,Z]) \ .
\end{array}
$
\caption{{\sf 
(a)
The space of local (affine) CY3 thus far charted.
The crosses symbolize isolated cases and the Venn diagram are the major families studied.
Here we have marked $\IC^3$ and the conifold $\cC$.
(b)
The clover quiver for $\cN=4$ super-Yang-Mills, corresponding to $\IC^3$.
}}
\label{f:CY3affine}
\end{center}
\end{figure}

As far as computational geometry is concerned, this amounts to 
\begin{quote}
{\bf Challenge:} A cartography of the space of non-compact, affine (and typically singular) CY3.
\end{quote}
Such CY3 are also called {\it local} and admit crepant resolutions to smooth CY3.
In complex dimension two, we know that CY2 is the K3 surface and locally these are the ADE surface singularities on which we will expound shortly.
Again, in dimension 3, we are in {\it terra incognita}.
In Figure \ref{f:CY3affine} we draw the counterpart to Figure \ref{f:CY3} and present a topologically correct and metrically irrelevant Venn diagram of the space of affine CY3.
Indeed, here we have several infinite families.

The simplest affine CY3 is, of course, trivially $\IC^3$, which is utterly, and not merely Ricci, flat.
The dual gauge theory is the famous $\cN=4$ super-Yang-Mills and the SE base, simply $S^5$ over which $\IC^3$ is a cone; this was Maldacena's archetypal case.
In the 15 years since 1998, in generalizing this, a tremendous amount of new physics and mathematics has emerged. We will now take a glimpse from the perspective of algorithmic geometry, the multitude of progenitors of this theory.
This {\it parent} theory is best described as a {\bf quiver} which is a finite directed graph whose nodes are factors in a product gauge group, usually taken to be $\prod\limits_i SU(N_i)$ and whose arrows from node $i$ to $j$ are bi-fundamentals $(\fund,\antifund)$ of $SU(N_i) \times SU(N_j)$ (those from nodes to themselves are adjoints). Finally, closed loops in the quiver, formed by tracing along directed paths, are gauge invariant operators and could be terms in the superpotential.

The $\IC^3$ theory is conveniently summarized as a {\it clover quiver} with the single node representing $U(N)$ and the three edges, the three adjoint fields $X,Y,Z$. Moreover, there is a standard cubic superpotential $W = \tr(X[Y,Z])$.
We present this in part (b) of Figure \ref{f:CY3affine}.
Our emphasis in this section will not so much be on breaking this $\cN=4$, $U(n)$ gauge theory to Standard-like models with $\cN=1$ SUSY, which is itself an extensive subject \cite{Uranga:2007zza,Verlinde:2005jr,Gmeiner:2005vz,Antoniadis:2002qm,Berenstein:2001nk,Bailin:2000kd,Aldazabal:2000sa}, but more on the wealth of geometrical methods which have arisen, the classifications which have been addressed, as well as the computational challenges ahead.
Thus the dialogue between the geometry of affine CY3 and the physics and combinatorics of quivers will be our focal point.

\subsection{Orbifolds}\label{s:orb}
The simplest class of affine CY3 is clearly (Gorenstein) quotients of $\IC^3$; these are the {\it orbifolds} of the form $\IC^3 / \Gamma$ where $\Gamma$ is a discrete finite subgroup of $SU(3)$.
We can obtain the quiver from the parent $\IC^3$ by $\Gamma$-projection \cite{Lawrence:1998ja,Kachru:1998ys,Johnson:1996py,He:2004rn}.
Let $\{ {\bf r}_i \}$ be the set of irreducible representations of $\Gamma$ and ${\cal R}$ a chosen representation (for fermions, this is the $4 = 1\oplus1\oplus2$ and for bosons, the $6 =  1\oplus1\oplus2\oplus2$, coming from the $SU(4)$ R-symmetry of the parent $\cN=4$). Next, form the decomposition
\begin{equation}\label{adjmat}
{\cal R}\otimes {\bf r}_i = \bigoplus\limits_{j}a_{ij}^{{\cal R}} {\bf r}_j \ .
\end{equation}
The resulting theory is the quiver whose {\it adjacency matrix} is given by $a_{ij}^{{\cal R}}$.
Explicitly, we can invert the above using characters to obtain
$a_{ij}^{{\cal R}}=\frac{1}{\left| \Gamma \right|}\sum\limits_{\gamma
=1}^{r}r_{\gamma }\chi_{\gamma }^{{\cal R}}\chi_{\gamma }^{(i)}\chi
_{\gamma }^{(j)*}$, where $r_{\gamma }$ is the order of the conjugacy class
containing $\gamma$ and $\chi_{\gamma }^i$ is the character of
$\gamma$ in the $i$-th representation.

The situation in complex dimension 2 is familiar to algebraic geometers.
As mentioned above, local CY2 - i.e., K3 surfaces - were already classified in the beginning of C20th, and fall under an ADE pattern \cite{duVal}.
Explicitly, the affine equations, as hypersurface singularities in $\IC[x,y,z]$, are
\begin{equation}
\begin{array}{cc}
\begin{array}{l}\includegraphics[scale=0.4]{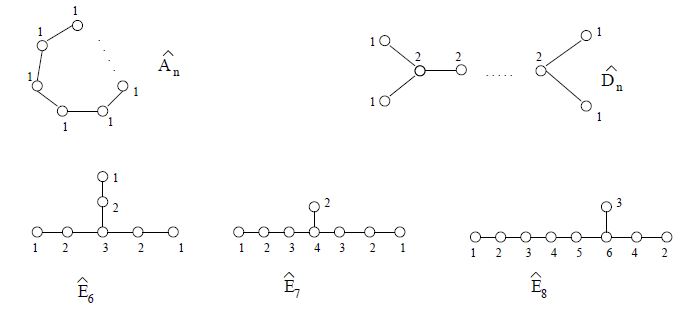}\end{array}
&
\begin{array}{ll}
A_n: &xy+z^n=0\\
D_n: &x^2+y^2z+z^{n-1}=0\\
E_6: &x^2+y^3+z^4=0\\
E_7: &x^2+y^3+yz^3=0\\
E_8: &x^2+y^3+z^5=0 \ ,
\end{array}
\end{array}
\end{equation}
corresponding to the orbifold $\IC^2 / \Gamma$ with $\Gamma$ discrete finite subgroups of $SU(2)$, which are the cyclic ($\hat{A}_n \sim \IZ_{n+1}$), binary dihedral ($\hat{D}_n$) and binary exceptional ($\hat{E}_{6,7,8}$) groups.
The celebrated result of McKay \cite{mckay} states that the adjacency matrices $a_{ij}$ in \eqref{adjmat} are precisely the associated affine Dynkin diagrams.
Thus our quiver gauge theories furnish an elegant physical realization of the McKay Correspondence \footnote{
Over the past couple of years wherein I have had the fortune to consolidate my friendship with John, who like a grand-father calls me on skype almost daily to chat on mathematics and life, I had the singular opportunity to witness the alertness of his mind and the breadth of his knowledge.
Working with him is rather like the challenge and the enjoyment of reading Joyce's {\it Ulysses}, one is transported to a cosmos most intricate and vast, filled with connexions and allusions, infused with amusements and humour, and one is always stricken by wonder.
}.

Our present situation of dimension 3 follows a similar pattern
(we need to point out that unlike dimension 2, the uniqueness and existence of crepant resolutions is not guaranteed here.
).
Other than the obvious $\IZ_k \times \IZ_{k'}$, the non-Abelian subgroups are \cite{blich}
\begin{equation}
\begin{array}{|c|c|}\hline
\mbox{Infinite Series} & \Delta(3n^2), \Delta(6n^2) \\ \hline
\mbox{Exceptionals} & 
\Sigma_{36 \times 3},
\Sigma_{60 \times 3},
\Sigma_{168 \times 3},
\Sigma_{216 \times 3}, \Sigma_{360 \times 3} \\
\hline\end{array}
\end{equation}
The quivers for these, using \eqref{adjmat}, were painstakingly catalogued \cite{Hanany:1998sd}, especially with the aid of computer systems \cite{GAP} and many interesting structures can be uncovered.
One could proceed further and ask for the full $SU(4)$ group. Here, supersymmetry will be broken so the physics is not as well controlled. Nevertheless, the full set of group and quivers can be calculated, combining classical algebra and modern computing \cite{blich,Hanany:1999sp,GAP}.

\subsection{del Pezzo}\label{s:dP}
A natural way to construct affine CY3, we have seen, is to realize it as a real cone over a smooth SE 5-fold.
There is a complex analogue of this whereby one realizes the CY3 $X$ as a complex cone over some complex surface $S$.
Indeed, when the SE is regular, itself can be realized as a $U(1)$ bundle over $S$.
A simple solution is to have $S$ possess appropriate positive curvature so that the cone metric ``cancels'' to give overall zero curvature for $X$.
We have seen the analogue of this in the compact situation in \S\ref{s:elliptic}, where the CY3 is an elliptic fibration over some base so that the overall first Chern class vanishes.
Complex (K\"ahler) manifolds admitting positive Ricci curvature are called {\bf Fano} and in complex dimension 2, they are Fano surfaces \cite{duca} with ample anti-canonical bundle.
These surfaces are the {\bf del Pezzo} surfaces which are simply $\IP^2$ blown-up at $k=0,\ldots,9$ {\it generic} points, denoted as $dP_k$, and the zeroth {\bf Hirzebruch surface} $F_0 = \IP^1 \times \IP^1$.
The family tree is:
\begin{equation}\label{dP}
\begin{array}{ccc}
&(F_0 = \IP^1 \times \IP^1) &\\
&\downarrow&\\
(dP_0 = \IP^2) \rightarrow &dP_1 &\rightarrow dP_2 \rightarrow \ldots
\rightarrow dP_8 \rightarrow (dP_9 = \frac12 K3) \ ,
\end{array}
\end{equation}
where an arrow denotes a blowup by $\IP^1$.
The extremal case of $k=9$ is usually called half-K3 since its first Chern class squares to 0.
Moreover, one could fathom blowing up $F_0$ at various generic points, however, at one point blown-up, the result is already isomorphic (bi-rational) to $dP_1$
and thus the families converge and no new progeny is produced.

For reference, the non-trivial homology (curve classes) is
\begin{equation}\begin{array}{rcl}
H_2(dP_k; \IZ) &=&  \gen{\ell, E_{i=1,\ldots,k} | \ell^2 =1, \ \ell \cdot E_i = 0, \ E_i \cdot E_j=- \delta_{ij}} \ ; \\
H_2(F_0; \IZ) &=& \gen{S,E | E^2 = S^2 = 0, \ S \cdot E = 1} \ ,
\end{array}
\end{equation}
where, clearly, for $dP_k$, $\ell$ is the class of the $\IP^1 \subset \IP^2$ and $E_i$ are the (exceptional) $\IP^1$-blowups and for $F_0$, $S$ and $E$ are the two $\IP^1$s. Furthermore, the Chern classes are
$c_1(dP_k) = 3\ell - \sum\limits_{i=1}^k E_i \ , c_2(dP_k) = 3+k$ and
$c_1(F_0) = 2S+2E \ , c_2(F_0) = 4$.

The astute reader would recognize $H_2(dP_k; \IZ)$ as the root lattice of the exceptional Lie algebra $\IE_k$.
This Mckay-esque curiosity was further explored \cite{Iqbal:2001ye}, wherein the remarkable observation that \eqref{dP} resembles the structure of M-theory compactification was made.

Explicit equations for these surfaces can be written as projective varieties (for example, $dP_k$ is of degree $9-k$), with rather complicated equations.
Famous is $dP_6$; this can be realized as the cubic (degree $9-6=3$)  surface in $\IP^3$, a classical object know to the C19th.
To obtain the CY3 cone, one simply de-homogenizes and writes these as affine equations.

How does one geometrically engineer the gauge theory?
It turns out that $F_0$ and $dP_{k \le 3}$ afford toric description, the details of which we shall shortly visit.
In general one could make use of so-called exceptional collections of bundles to compute the quiver and superpotential \cite{Herzog:2003dj,Herzog:2003zc,Herzog:2005sy,Hanany:2006nm,Aspinwall:2004vm,Wijnholt:2002qz}, which can be further exploited for MSSM model building
\cite{Dolan:2011qu,Cicoli:2012vw,Malyshev:2007yb,Verlinde:2005jr}.
One structure of note is that these del Pezzo quivers organize themselves into ``blocks'' wherein three groups (blocks) of nodes suffice to exhibit the symmetry \cite{KN,Benvenuti:2004dw,Hanany:2012mb}.

\subsection{Toric CY3}
We have reserved, as in the compact case in the previous section, the largest dataset for the last. These are the toric CY3 spaces. 
Of course, we need to emphasize there are no compact CY3s which are toric (cf.~a nice introduction \cite{Bouchard:2007ik}).
As much as orbifolds reduce geometry to finite group representations, toric geometry reduces the CY3 to the investigation of the combinatorics of integer cones. 

A few points deserve emphasis.
Because our CY3 is affine and local, we need not consider the glueing of cones into fans as is done for the standard compact toric variety (such as the ambient fourfolds in the KS dataset). Thus our situation is easier and each of our singular CY3 is described by a single integer convex cone in $\IZ^3$.
The Calabi-Yau condition implies that the endpoints of the integer vectors be co-planar. Thus we have the remarkably simple description for each toric CY3: a (convex) grid of integer points in $\IZ^2$.

All $SU(3)$ Abelian orbifolds (including the parent herself) of $\IC^3$ are toric CY3. The cone for $\IC^3$ is generated by the 3 standard basis vectors $e_{1,2,3}$ in $\IZ^3$, which, upon exponentiation by the coordinates, give respectively the three monomial generators $x,y,z$. And indeed these are free generators without relations: ${\rm Spec}\IC[x,y,z] \simeq \IC^3$, as required.
Thus the {\em toric diagram} consists of the endpoints of these three standard basis vectors, which are indeed co-planar and can thus, after appropriate linear fractional transformation, be chosen to be the three lattice points: $\{(0,0), (0,1), (1,0)\} \subset \IZ^2$.

\begin{figure}[t]
\begin{center}
\includegraphics[scale=0.25]{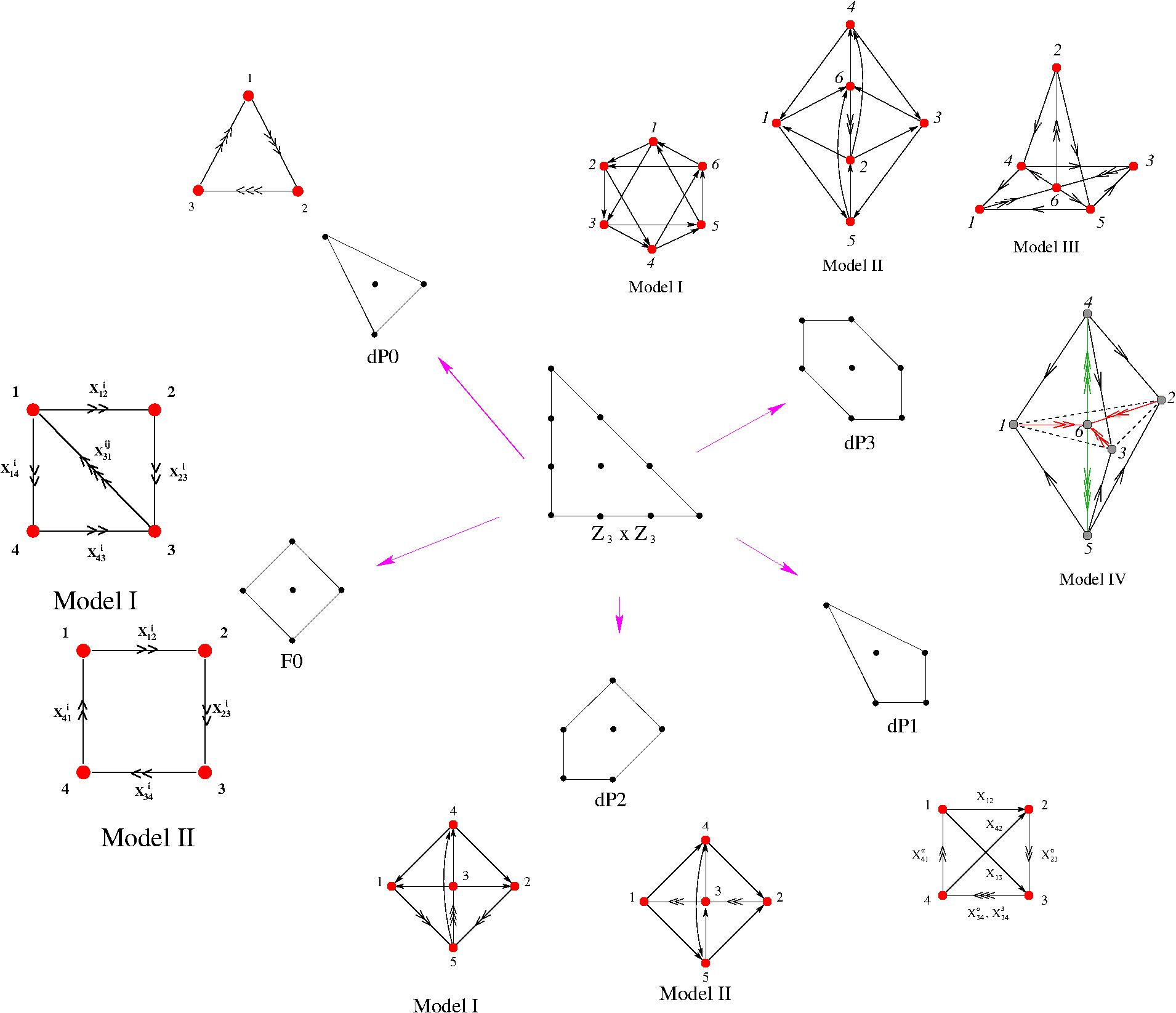}
\caption{{\sf 
The toric diagrams of some partial resolutions (sub-diagrams) of the Abelian orbifold $\IC^3/\IZ_3^2$, as well as the quivers for the engineered 4d $\cN=1$ gauge theories.
The various different phases corresponding to the same geometry are Seiberg duals.
}}
\label{f:c3z3z3}
\end{center}
\end{figure}

As mentioned in \S\ref{s:orb}, the Abelian CY3 orbifolds are of the form $\IZ_m \times \IZ_n$. The toric diagram for these are simply the enlargement of the triangle for $\IC^3$ into an $m \times n$ triangle, including all interior and boundary lattice points in $\IZ^2$.
The key point is that {\em any} CY3 toric diagram is a (convex) sub-diagram of this for sufficiently large $m$ and $n$.
In Figure \ref{f:c3z3z3} we show the toric diagram of $\IC^3 / \IZ_3^2$ and its various sub-diagrams.
Reducing to a sub-diagram is called {\em partial resolution}.

Using Witten's gauged linear sigma model \cite{Witten:1993yc}, the technique of obtaining the gauge theory of the D-brane probe on toric CY3 was developed \cite{Douglas:1997de,Beasley:1999uz} and algorithmized \cite{Feng:2000mi}.
The partial resolution corresponds to Higgsing in the gauge theory and since the quiver and superpotential for the Abelian orbifold can be constructed readily using the methods in \S\ref{s:orb}, the problem of constructing the general dual gauge theory is reduced to the combinatorics of systematically reducing nodes from the toric diagram.
This was the state of the art for about a decade, the only hurdle being the exponential growth-rate in complexity as the number of lattice points increases.
Nevertheless, a wealth of gauge theories were obtained \cite{Feng:2000mi,Feng:2001xr,Feng:2001bn,Benvenuti:2004dy,Gauntlett:2004hh,Beasley:2001zp}.
In Figure \ref{f:c3z3z3} we also include some quivers for reference.

Given the cumulating data, a few observations could be made for these toric gauge theories, some by construction, some by geometrical engineering and some empirical and remained mysterious for a while.
These came to be known as the {\em toric conditions}:
\begin{itemize}
\item All nodes of the quiver are rank 1, i.e., we have a $U(1)^k$ gauge theory \footnote{Of course, we can augment the gauge group, by a stack of $N$ branes, to $U(N)^k$, whereby the fields are promoted from complex numbers to matrices; nevertheless, unlike the orbifold or general del Pezzo cases, we do not automatically have the freedom of unequal ranks.}; this is due to the fact that underlying toric varieties are $\IC^*$-actions.
\item All fields (arrows in the quiver) appear in the superpotential exactly twice with opposite sign; this is due to the so-called {\it binomial ideal} definition of a toric variety \cite{binomial}.
\item Let there be $N_G$ gauge group factors, $N_E$ fields and $N_W$ terms in the superpotential; then, curiously, $N_G - N_E + N_W = 0$.
For example for the famous $N=4$ SYM, there $N_G=1$, $N_E=3$ and $N_W=2$.
\end{itemize}

Furthermore, the attentive reader would see that the mapping from the quiver gauge theory to the toric diagram is often many to one, i.e., there are often several physical theories sharing the same moduli space in the IR.
Over the years, this was realized to be Seiberg duality for $\cN=1$ gauge theory \cite{Feng:2001bn,Beasley:2001zp,Cachazo:2001gh,Cachazo:2001sg,Franco:2004jz}.
There are two extraordinary facts about this enormous subject.
The operation on the dual quivers has been known independently in the mathematical community as {\it mutation} \cite{FZ}.
Recently, this duality move has been realized in yet another fundamental and seemingly unrelated field, that of computation of on-shell amplitudes in SYM \cite{ArkaniHamed:2012nw}.

The situation drastically changed in 2005 when it was realized that all these toric gauge theories can be completely encoded by a bipartite graph drawn on a torus, or equivalently a doubly periodic tiling of the plane, known as a dimer model or brane tiling \cite{Hanany:2005ve}.
Much activity ensued \cite{Franco:2005rj,GarciaEtxebarria:2006aq,Franco:2005sm,Feng:2005gw} and by now it is clear that the dimer description of quiver gauge theories on toric CY3 is the most conducive and enlightening one.
The material has blossomed substantially and the reader is referred to two excellent reviews \cite{Kennaway:2007tq,Yamazaki:2008bt} as well as a rapid introduction \cite{He:2012js}.
We remark that the origin of the topological relation in the third of the toric condition is conformality, while that of the bipartite-ness is geometrical, viz., the binomial ideal definition of a toric variety.
It is intriguing that from such seemingly technical definition could stem so much physics.

Once we are in the world of embedded bipartite graphs on Riemann surfaces, it is inevitable that we touch upon Grothendieck's {\it dessin d'Enfant}, and thence, aspects of algebraic number theory.
Let us end this subsection on toric gauge theories with a view toward dessins \cite{He:2012kw,Hanany:2011ra,Jejjala:2010vb,Ashok:2006br}.
The relation between dessins and elliptic curves (CY1) as well as K3 surfaces (CY2) have been established over the years, it seems that the correspondence persists to our present case of dimension 3, and possibly beyond.

\subsection{The Plethystic Programme}
As a parting topic in our geometry-gauge theory correspondence for affine CY3, let us discuss the important matter of enumeration of operators.
In the spirit of the super-conformal index \cite{Witten:2000nv}, extensively studied \cite{Romelsberger:2005eg,Bhattachar ya:2008zy,Gadde:2011uv}, a so-called {\it Plethystic Programme} was introduced to study general gauge theories with supersymmetry, especially those with non-trivial VMS \cite{Benvenuti:2006qr,Feng:2007ur}.
The methods thus readily adapt to our case of VMS being affine CY3 and the point d'appui is an object familiar to classical algebraic geometry, viz., the {\em Hilbert series}, the calculation of which has also recently been of algorithmic interest \cite{m2,sing,bertini}.

The programme proceeds with the following algorithm:
\begin{itemize}
\item Find the vacuum geometry $\cM$ of the theory, which is the algebraic variety parametrized by the vacuum expectation values of the scalars. Compute the Hilbert series 
\begin{equation}
f(t) = \sum\limits_{n=0}^\infty a_n t^n \ , \qquad a_n \in \IZ_{\ge 0}
\end{equation}
of $\cM$ with respect to some appropriate grading dictated by the natural charges in the system. This is the generating function for counting the basic single-trace invariants.
For example, the Hilbert series of $\IC^3$ is $f(t) = (1-t)^{-3}$, thus the Taylor coefficient $a_n = \frac12 (n+2)(n+1)$ is the number of single-trace 1/2-BPS operators at R-charge $n$.

\item To find the multi-trace objects, i.e., the unordered products of the single-traces, we take the plethystic exponential (sometimes known as the Euler transform)
\begin{equation}
g(t) =  PE[f(t)] := \exp\left( \sum_{n=1}^\infty \frac{f(t^n) -
  f(0)}{n} \right) = {\prod\limits_{n=1}^\infty (1-t^n)^{-a_n}}  \ .
  \end{equation}

\item 
There is an analytic inverse function to $PE$, which is the plethystic logarithm, given by
\begin{equation}
f(t) = PE^{-1}(g(t)) = \sum_{k=1}^\infty
\frac{\mu(k)}{k} \log (g(t^k)) \ , 
\end{equation}
where $\mu(k)$ is the M\"obius function
\[
\mu(k) := \left\{\begin{array}{lcl}
0 & & k \mbox{ has repeated prime factors}\\
1 & & k = 1\\
(-1)^n & & k \mbox{ is a product of $n$ distinct primes} \ .
\end{array}\right.
\] 
The plethystic logarithm of the Hilbert series gives the geometry of $\cM$, i.e., 
\[
PE^{-1}[f(t)] = \mbox{ defining equation of $\cM$}.
\] 
In particular, if $\cM$ were complete intersection, then $PE^{-1}[f(t)]$ is polynomial.

\item The Hilbert series of the $N$-th symmetric product is given by
\begin{equation}
g_N(t; \cM) = f(t;{\rm sym}^N(\cM)), \qquad {\rm sym}^N(\cM) := \cM^N/S_N \ ,
\end{equation}
where the ``grand-canonical'' partition function is given by the fugacity-inserted plethystic exponential of the Hilbert series:
\begin{equation}
g(\nu ; t) = PE_\nu[f(t)] := \prod\limits_{n=0}^{\infty} 
{(1 - \nu  \, t^n)^{-a_n}} = \sum\limits_{N=0}^\infty g_N(t) \nu^N  \ .
\end{equation}
In the gauge theory, this is considered to be at finite $N$ and the expansion
$g_N(t)  = \sum\limits_{n=0}^\infty b_n t^n$ gives the number $b_n$ of operators of charge $n$.
\end{itemize}
One very practical aspect of the plethystic programme is that it could test certain geometrical properties of the VMS.
Suppose one has the explicit polynomial ideal describing $\cM$, then a wonderful theorem of Stanley \cite{stanley,Forcella:2008bb} dictates that if the Hilbert series has a palindromic numerator, then $\cM$ is Calabi-Yau.

\section{Vacuum Geometry: Search for New Signatures}
From the standpoint of computational algebraic geometry, we have described the various popular databases of compact and non-compact CY3 in the previous two sections.
From the perspective of phenomenology, particularly string phenomenology, we can regard the aforementioned as a ``top-down'' approach.
Indeed, the plethora of CY3 is part of the {\it vacuum degeneracy problem} where an overwhelming number of geometries seems to be candidates in giving Standard-like low-energy behaviour.
Of course, the lesson is the word ``like'': we have seen over the last decade or so that even the mildest constraint such as having exact particle content already cuts potential candidates by a factor of billions.

In some sense, string phenomenology, due to the rapid advance in mathematics and computation, has reached a stage akin to the hunt for exo-planets.
While large scale telescopes operating at a wide range of wave-lengths are sifting the visible universe for planet similar to our own, the string vacuum project uses modern computing to sift through Calabi-Yau and other geometries whose compactifications give universes with {\em our} Standard Model.

Can one take a complementary view?
Inspired by the study of vacuum moduli spaces, wherein a natural (complex) geometry is associated to a (supersymmetric) gauge theory, and aided by the techniques from the plethystic programme, let us forget about Calabi-Yau manifolds and about string theory and step back to purely consider field theory.
Surprisingly, we will find that we are compelled to return to the Calabi-Yau world.

Now, for an arbitrary $\cN=1$ four-dimensional gauge theory, we can find its (classical) VMS by computing the F-terms and D-terms.
With the sophistry of symplectic geometry, this has been rephrased as the geometric invariant theory (GIT) quotient of the space of solutions of the F-terms prescribed by the Jacobian of the holomorphic superpotential by the gauge fixing conditions provided by the D-terms.
From the point of view of algorithmic geometry \cite{Gray:2006jb,Gray:2009fy,Hauenstein:2012xs}, this is an elimination problem in polynomial ideals:
\begin{itemize}
\item INPUT:
\begin{enumerate}
\item Superpotential $W(\{\phi_i\})$, a polynomial in variables $\phi_{i=1, \ldots, n}$, corresponding to the vacuum expectation values of the scalar fields in the chiral multiplet, charged as adjoints, bi-fundamentals and more complicated representations of some product gauge group;
\item Generators of gauge invariants: $r_j(\phi_i)$, $j=1, \ldots, k$ polynomials in $\phi_i$, these are primitive single-trace operators in the fields. For quiver theories, these are minimal loops in the directed graph, where composition of arrows are matrix multiplication according to the ranks of the nodes;
\end{enumerate}

\item ALGORITHM:
\begin{enumerate}
\item Define the polynomial ring $R = \IC[\phi_{i=1,\ldots,n}, y_{j=1,\ldots,k}]$,
\item Consider the ideal $I = \gen{\frac{\partial {W}}{\partial \phi_i}; y_j - r_j(\phi_i)}$,
\item Eliminate all variables $\phi_i$ from $I \subset R$, giving the ideal $\cM$ in terms of $y_j$;
\end{enumerate}

\item OUTPUT:
$\cM$ corresponds to the VMS as an affine variety in $\IC[y_1, \ldots, y_k]$.
\end{itemize}

This computation is perfectly adapted for a Gr\"obner basis treatment (since, afterall, the latter is a polynomial generalization of Gaussian elimination \cite{Gray:2009fy}) and many freely available software have been tailored \cite{m2,sing,sage,Gray:2008zs,Gray:2007yq}; of these we have taken liberal advantage over the years and have been able to distill much useful information.
The draw-back is that Gr\"obner basis computations suffer from exponential growth in memory usage and running time as well as non-parallelizability.
Recently \cite{Mehta:2012wk,Hauenstein:2012xs,He:2013yk}, it was realized that if all that is needed are basic geometrical data such as dimension, degree and Hilbert series, then it suffices to use so-called homotopy continuation methods in {\it numerical algebraic geometry} which are, crucially, parallelizable.
Again, there is publicly available software to our ready assistance \cite{bertini}.

\subsection{The Geometry of the Standard Model}
Thus armed, mathematically and computationally, the most natural question to ask would be: what is the underlying geometry of the most important gauge theory of them all, the (supersymmetric) Standard Model \footnote{An analogue situation for the regular non-SUSY Standard Model has also been considered, wherein the flavour invariants were studied \cite{Hanany:2010vu}}?
The first steps toward answering this question have been addressed \cite{Gray:2005sr,Gray:2006jb}.

\def\eps{\epsilon^{\alpha \beta}}
\def\barH{\overline{H}}

To give an idea of the complexity of the input data, let us consult once more the list of fields from \eqref{mssmfields}: $Q_{a, \alpha}^i$, the $SU(2)_L$ doublet quarks; $u_a^i$, the $SU(2)_L$ singlet up-quarks; $d_a^i$, the $SU(2)_L$ singlet down-quarks; $L_{\alpha}^i$, the $SU(2)_L$ doublet leptons; $e^i$, the $SU(2)_L$ singlet leptons, as well as  $H_\alpha$, the up-Higgs and $\barH_\alpha$, the down-Higgs, with indices $i,j,k,l = 1,2,3$ (Flavour), $a,b,c, d = 1,2,3$ ($SU(3)_C$ colour) and $\alpha, \beta, \gamma, \delta = 1,2 $ ($SU(2)_L$-indices), giving us a total of $18+9+9+6+3+2+2 = 49$ scalar components.
The minimal renormalizable superpotential is
\begin{align}
W_{\rm minimal} &= C^0 \sum_{\alpha, \beta} H_\alpha \barH_\beta \eps + \sum_{i,j} C^1_{ij} \sum_{\alpha, \beta, a} Q^i_{a,\alpha} u^j_a H_\beta \eps \nonumber \\
& +\sum_{i,j} C^2_{ij} \sum_{\alpha, \beta, a} Q^i_{a,\alpha} d^j_a \barH_\beta \eps + \sum_{i,j} C^3_{ij} e^i \sum_{\alpha, \beta} L^j_{\alpha} \barH_\beta \eps \ .
\end{align}

The true bottle-neck, however, is the minimal set of gauge invariants operators; which has nevertheless been known for some time \cite{Gherghetta:1995dv}:
\[
\mbox{
{\scriptsize
\begin{tabular}{|c||c|c|c|}\hline
\mbox{Type} & \mbox{Explicit Sum} & \mbox{Index} & \mbox{Number} \\
\hline \hline
$LH$  & $L^i_\alpha H_\beta \eps$ & $i=1,2,3$ & 3 \\ \hline
$H\barH$ & $H_\alpha \barH_\beta \eps$ & & 1  \\ \hline
$udd$ & $u^i_a d^j_b d^k_c \epsilon^{abc}$ & $i,j=1,2,3$; $k=1,\ldots,j-1$ & 9  
\\ \hline
$LLe$ & $L^i_\alpha L^j_\beta e^k \eps$ & $i,j=1,2,3$; $k=1,\ldots,j-1$ & 9  \\ 
\hline
$QdL$ & $Q^i_{a, \alpha} d^j_a L^k_\beta \eps$ & $i,j,k=1,2,3$ & 27 \\ \hline
$QuH$ & $Q^i_{a, \alpha} u^j_a H_\beta \eps$ & $i,j=1,2,3$ & 9 \\ \hline
$Qd\barH$ & $Q^i_{a, \alpha} d^j_a \barH_\beta \eps$ & $i,j=1,2,3$ & 9 \\ \hline
$L\barH e$ & $L^i_\alpha \barH_\beta \eps e^j$ & $i,j=1,2,3$ & 9 \\ \hline
$QQQL$ & $Q^i_{a, \beta} Q^j_{b, \gamma} Q^k_{c, \alpha} L^l_\delta \epsilon^{ab
c} \epsilon^{\beta\gamma}\epsilon^{\alpha\delta}$ & 
$\begin{array}{l} i,j,k,l=1,2,3; i\ne 
k, j\ne k, \\ j<i, (i,j,k) \ne (3,2,1) \end{array}$ & 24 \\ \hline
$QuQd$ & $Q^i_{a, \alpha} u^j_a Q^k_{b, \beta} d^l_b \eps$ & $i,j,k,l=1,2,3$ & 8
1 \\ \hline
$QuLe$ & $Q^i_{a, \alpha} u^j_a L^k_{\beta} e^l \eps$ & $i,j,k,l=1,2,3$ & 81 \\ 
\hline
$uude$ & $u^i_a u^j_b d^k_c e^l \epsilon^{abc}$ & $i,j,k,l=1,2,3; j<i$ & 27 \\ \hline
$QQQ\barH$ & $Q^i_{a, \beta} Q^j_{b, \gamma} Q^k_{c, \alpha} \barH_\delta \epsilon^{abc} \epsilon^{\beta\gamma} \epsilon^{\alpha\delta}$ & 
$\begin{array}{l} i,j,k,l=1,2,3; i\ne k, j\ne k, \\ j<i, (i,j,k) \ne (3,2,1) \end{array}$ & 8 \\ \hline
$Qu\barH e$ & $Q^i_{a, \alpha} u^j_a \barH_\beta e^k \eps$ & $i,j,k =1,2,3$ & 27 \\ \hline
$dddLL$ & $d^i_a d^j_b d^k_c L^m_\alpha L^n_\beta \epsilon^{abc} \epsilon_{ijk} \eps$ & $m,n=1,2,3, n<m$ & 3 \\ \hline
$uuuee$ & $u^i_a u^j_b u^k_c e^m e^n \epsilon^{abc} \epsilon_{ijk}$ & $m,n=1,2,3, n \le m$ & 6 \\ \hline
$QuQue$ & $Q^i_{a, \alpha} u^j_a Q^k_{b, \beta} u^m_b e^n \eps$ & $ \begin{array}{l} i,j,k,m,n=1,2,3; \\ \mbox{antisymmetric}\{ (i,j), (k,m) \}
\end{array}$ & 108  \\ \hline
$QQQQu$ & $Q^i_{a, \beta} Q^j_{b, \gamma} Q^k_{c, \alpha} Q^m_{f,\delta} u^n_f \epsilon^{abc} \epsilon^{\beta\gamma} \epsilon^{\alpha\delta}$ & $\begin{array}{l} i,j,k,m=1,2,3; i\ne m, j\ne m, \\ j<i, (i,j,k) \ne (3,2,1) \end{array}$ & 72  \\ \hline
$dddL\barH$ & $d^i_a d^j_b d^k_c L^m_\alpha \barH_{\beta} \epsilon^{abc}\epsilon_{ijk} \eps$ & $m=1,2,3$ & 3 \\ \hline
$uudQdH$ & $u^i_a u^j_b d^k_c Q^m_{f, \alpha}d^n_f H_\beta \epsilon^{abc} \eps$ & $i,j,k,m=1,2,3; j<i$ & 81 \\ \hline
$(QQQ)_4LLH$  & $(QQQ)_4^{\alpha\beta\gamma} L^m_\alpha L^n_\beta H_\gamma$ & $m,n=1,2,3, n<=m$ & 6 \\ \hline
$(QQQ)_4LH\barH$ & $(QQQ)_4^{\alpha\beta\gamma} L^m_\alpha H_\beta \barH_\gamma$ & $m=1,2,3$ & 3 \\ \hline
$(QQQ)_4H\barH\barH$ & $(QQQ)_4^{\alpha\beta\gamma} H_\alpha \barH_\beta \barH_\gamma$ & & 1 \\ \hline
$(QQQ)_4LLLe$ & $(QQQ)_4^{\alpha\beta\gamma} L^m_\alpha L^n_\beta L^p_\gamma e^q$ & $m,n,p,q=1,2,3, n\le m,p \le n$ & 27 \\ \hline
$uudQdQd$ & $u^i_a u^j_b d^k_c Q^m_{f, \alpha}d^n_f Q^p_{g, \beta} d^q_g \epsilon^{abc} \eps$ & $\begin{array}{l} i,j,k,m,n,p,q=1,2,3; \\ j<i, \mbox{antisymmetric}\{(m,n), (p,q)\} \end{array}$ & 324 \\ \hline
$(QQQ)_4LL\barH e$ & $(QQQ)_4^{\alpha\beta\gamma} L^m_\alpha L^n_\beta \barH_\gamma e^p$ & $m,n,p = 1,2,3, n \le m$ & 9 \\ \hline
$(QQQ)_4L\barH\barH e$ & $(QQQ)_4^{\alpha\beta\gamma} L^m_\alpha \barH_\beta \barH_\gamma e^n$ & $m,n=1,2,3$ &  9 \\ \hline
$(QQQ)_4\barH\barH\barH e$ & $(QQQ)_4^{\alpha\beta\gamma} \barH_\alpha \barH_\beta \barH_\gamma e^m$ & $m=1,2,3$ &  3 \\ \hline
\end{tabular}
}
}
\]
There are 28 types as above, totalling 991 invariants.
Thus we need to work over a polynomial ring in $991 + 49 = 1040$ complex variables, a tremendous task indeed.
Clearly, this is beyond any conceivable Gr\"obner basis analysis.
However, with enough computing power, it could be within the parallelizable scope of numerical algebraic geometry, and is currently being set up.

As bench-marks, one could enquire about two limits: the EW sector and the pure QCD sector.
Here, we find rather intriguing geometries.
Setting the quarks to zero (and not worrying about anomaly for the moment), we find the VMS is three complex dimensional which explicitly is an affine cone over a classical object, viz., the Veronese surface.
This is a curious appearance of this geometry \cite{Gray:2005sr,Gray:2006jb}.

Now, for the quark sector, the situation is truly remarkable.
For pure sQCD with $N_f$ flavours and $N_c$ colours, and no superpotential, it is a standard fact that the VMS is of dimension 
$N_f^2$ for $N_f < N_c$ and
$2N_c N_f - (N_c^2 - 1)$  for $N_f  \ge  N_c$.
With our technology, we can actually find out what it is as an affine algebraic variety; it transpires that it is \cite{Gray:2008yu,Hanany:2008sb,Basor:2011da} Calabi-Yau!
Somehow, ``Calabi-Yau-ness'' is built into the very fabric of gauge theory.

~\\

With this tantalizing observation let us now take pause.
We have amused ourselves with a promenade in the land of Calabi-Yau threefold geometries, both the compact and the non-compact cases, and have witnessed that over the intervening years since the 1990's, the ``bestiary'' of Calabi-Yau manifolds \cite{hubsch} has grown to a rich and diverse kingdom.
Central to this explosion of information has been the rapid development of algorithmic geometry, powerful computing, and the ever-increasing potency of the cross-fertilization between mathematics and physics.
These three decades have fed us with a cornucopia of new data, new physics and new mathematics, but our feast on Calabi-Yau geometries has only just begun.

~\\

\centerline{ \reflectbox{\ding{167}}---\ding{69}---\ding{167} }

\newpage

\section*{Acknowledgments}
{\it Catharinae Sanctae Alexandriae et ad majorem Dei Gloriam.}

I would like to take this opportunity to thank my many friends and colleagues with whom I have enjoyed countless happy hours in working on the various subjects touched upon here, as well as the gracious patronage of the corresponding funding bodies:
\begin{itemize}
\item The Oxford-Philadelphia collaboration: Lara Anderson, Volker Braun, Philip Candelas, Xenia de la Ossa, Ron Donagi, Maxime Gabella, Seung-Joo Lee, Andre Lukas, Burt Ovrut, Tony Pantev, Rene Reinbacher, James Sparks, Balazs Szendroi (FitzJames Fellowship, Merton College, Oxford and UK STFC Advanced Fellowship);
\item The London collaboration: Amihay Hanany, Noppadol Mekareeya, Jurgis Pasukonis, Sanjaye Ramgoolam, Diego Rodriguez-Gomez, Rak-Kyeong Seong, Spyros Sypsas (UK STFC grant ST/J00037X/1);
\item The Boston-Edinburgh-London-Johannesburg-Munich collaboration: Ross Altmann, James Gray, Vishnu Jejjala, Benjamin Jurke, Brent Nelson, Joan Simon
(US NSF grant CCF-1048082);
\item The London-Rayleigh collaboration: Jonathan Hauenstein, Dhagash Mehta (SIAM);
\item My long-term collaborators Bo Feng, Sebatian Franco, John McKay, as well as my sometime coworkers Davide Forcella, Dan Grayson, Chris Herzog, Anton Ilderton, Kris Kennaway, Max Kreuzer, Nikos Prezas, Cumrum Vafa, Angel Uranga, Johannes Walcher, Alberto Zaffaroni;
\item My talented students Sun Chuang as well as Nessi Benishti, Sownak Bose, Joe Hewlett, James Read, Alexandru Valeanu, Mark van Loon (Merton College and Dept.~of Theoretical Physics and Mathematical Institute, Oxford, and studentships from The Nuffield Foundation, IoP, and EPSRC);
\item The London-Oxford-Tianjin collaboration: Mo-Lin Ge, Xue-Qian Li (Chinese Ministry of Education for endowing a Chang-Jiang chair Professorship and the City of Tianjin for a Qian-Ren Scholarship).
\end{itemize}

\newpage

\oddsidemargin   0in  
\evensidemargin  0in  
\textwidth       6.5in  
\renewcommand{\baselinestretch}{-2}



\begin{thebibliography}{9}
{\footnotesize
\bibitem{Candelas:1985en} 
  P.~Candelas, G.~T.~Horowitz, A.~Strominger, E.~Witten,
  ``Vacuum Configurations for Superstrings,''
  Nucl.\ Phys.\ B {\bf 258}, 46 (1985).


\bibitem{sing} 
G.-M.~Greuel, G.~Pfister, H.~Sch\"onemann,
  \emph{Singular: a computer algebra system for polynomial
    computations}, Centre for Computer Algebra, University of
  Kaiserslautern (2001).  Available at {\tt http://www.singular.uni-kl.de/}

\bibitem{m2}
D.~G.~Grayson, M.~E.~Stillman, 
``Macaulay2, a software system for research in algebraic geometry, ''
available at http://www.math.uiuc.edu/Macaulay2/.

\bibitem{GAP}
  The GAP~Group, 
``GAP -- Groups, Algorithms, and Programming, Version 4.6.4''; 2013,
  \verb+(http://www.gap-system.org)+.

\bibitem{sage}
William A. Stein et al.,
``Sage Mathematics Software''
The Sage Development Team, \verb|http://www.sagemath.org|.\\
for toric CY3, cf.~A.~Novoseltsev and V.~Braun,\\
\verb|http://www.sagemath.org/doc/reference/schemes/sage/schemes/toric/variety.html|

\bibitem{bertini}
D.J. Bates, J.D. Hauenstein, A.J. Sommese, C.W. Wampler.,
\newblock Available at {\tt www.nd.edu/$\sim$sommese/bertini}.


\bibitem{comp-book}
Y.-H. He, P. Candelas, A. Hanany, A. Lukas, B. Ovrut, Ed.
\newblock {\em {Computational Algebraic Geometry in String, Gauge Theory}}.
\newblock {Special Issue, Advances in High Energy Physics, Hindawi
  publishing,}, 2012, doi:10.1155/2012/431898.



\bibitem{Greene:1986ar} 
  B.~R.~Greene, K.~H.~Kirklin, P.~J.~Miron, G.~G.~Ross,
  ``A Superstring Inspired Standard Model,''
  Phys.\ Lett.\ B {\bf 180}, 69 (1986).

\bibitem{Candelas:2007ac} 
  P.~Candelas, X.~de la Ossa, Y.~-H.~He, B.~Szendroi,
  ``Triadophilia: A Special Corner in the Landscape,''
  Adv.\ Theor.\ Math.\ Phys.\  {\bf 12}, 429 (2008)
  [arXiv:0706.3134 [hep-th]].
  (cf.~A.~Ananthaswamy, Special Report, {\it New Scientist}, Jan 5th, 2008.)

\bibitem{Slansky:1981yr} 
  R.~Slansky,
  ``Group Theory for Unified Model Building,''
  Phys.\ Rept.\  {\bf 79}, 1 (1981).


\bibitem{Faraggi:2000pv} 
  A.~E.~Faraggi, M.~Pospelov,
  ``Selfinteracting dark matter from the hidden heterotic string sector,''
  Astropart.\ Phys.\  {\bf 16}, 451 (2002)
  [hep-ph/0008223].

\bibitem{Braun:2006da} 
  V.~Braun, E.~I.~Buchbinder, B.~A.~Ovrut,
  ``Towards realizing dynamical SUSY breaking in heterotic model building,''
  JHEP {\bf 0610}, 041 (2006)
  [hep-th/0606241].

\bibitem{Braun:2013wr} 
  V.~Braun, Y.~-H.~He, B.~A.~Ovrut,
  ``Supersymmetric Hidden Sectors for Heterotic Standard Models,''
  arXiv:1301.6767 [hep-th].


\bibitem{Distler:1987ee} 
  J.~Distler, B.~R.~Greene,
  ``Aspects of (2,0) String Compactifications,''
  Nucl.\ Phys.\ B {\bf 304}, 1 (1988).

\bibitem{hubsch}
T.~Hubsch.,
``Calabi-Yau Manifolds: a Bestiary for Physicists,''
World Scientific, 1992. ISBN 10: 981021927X 

\bibitem{Blumenhagen:1995ew} 
  R.~Blumenhagen, R.~Schimmrigk, A.~Wisskirchen,
  ``The (0,2) exactly solvable structure of chiral rings, Landau-Ginzburg theories, and Calabi-Yau manifolds,''
  Nucl.\ Phys.\ B {\bf 461}, 460 (1996)
  [hep-th/9510055].


\bibitem{Friedman:1997ih} 
  R.~Friedman, J.~W.~Morgan, E.~Witten,
  ``Vector bundles over elliptic fibrations,''
  alg-geom/9709029.


\bibitem{Friedman:1997yq} 
  R.~Friedman, J.~Morgan, E.~Witten,
  ``Vector bundles and  F theory,''
  Commun.\ Math.\ Phys.\  {\bf 187}, 679 (1997)
  [hep-th/9701162].

\bibitem{Donagi:1999ez} 
  R.~Donagi, B.~A.~Ovrut, T.~Pantev, D.~Waldram,
  ``Standard models from heterotic M theory,''
  Adv.\ Theor.\ Math.\ Phys.\  {\bf 5}, 93 (2002)
  [hep-th/9912208].


\bibitem{Donagi:2000zs} 
  R.~Donagi, B.~A.~Ovrut, T.~Pantev, D.~Waldram,
  ``Standard model bundles,''
  Adv.\ Theor.\ Math.\ Phys.\  {\bf 5}, 563 (2002)
  [math/0008010 [math-ag]].


\bibitem{Donagi:2004ia} 
  R.~Donagi, Y.~-H.~He, B.~A.~Ovrut, R.~Reinbacher,
  ``The Particle spectrum of heterotic compactifications,''
  JHEP {\bf 0412}, 054 (2004)
  [hep-th/0405014].


\bibitem{Donagi:2004ub} 
  R.~Donagi, Y.~-H.~He, B.~A.~Ovrut, R.~Reinbacher,
  ``The Spectra of heterotic standard model vacua,''
  JHEP {\bf 0506}, 070 (2005)
  [hep-th/0411156].

\bibitem{Donagi:2004su} 
  R.~Donagi, Y.~-H.~He, B.~A.~Ovrut and R.~Reinbacher,
  ``Higgs doublets, split multiplets and heterotic SU(3)(C) x SU(2)(L) x U(1)(Y) spectra,''
  Phys.\ Lett.\ B {\bf 618}, 259 (2005)
  [hep-th/0409291].


\bibitem{Curio:2004pf} 
  G.~Curio,
  ``Standard model bundles of the heterotic string,''
  Int.\ J.\ Mod.\ Phys.\ A {\bf 21}, 1261 (2006)
  [hep-th/0412182].

\bibitem{Andreas:2006zs} 
  B.~Andreas, G.~Curio,
  ``Stable bundle extensions on elliptic Calabi-Yau threefolds,''
  J.\ Geom.\ Phys.\  {\bf 57}, 2249 (2007)
  [math/0611762 [math-ag]].

\bibitem{Blumenhagen:2005ga}
  R.~Blumenhagen, G.~Honecker, T.~Weigand,
  ``Loop-corrected compactifications of the heterotic string with line bundles,''
  JHEP {\bf 0506} (2005) 020
  [hep-th/0504232].


\bibitem{Braun:2005zv} 
  V.~Braun, Y.~-H.~He, B.~A.~Ovrut, T.~Pantev,
  ``Vector bundle extensions, sheaf cohomology, and the heterotic standard model,''
  Adv.\ Theor.\ Math.\ Phys.\  {\bf 10}, 525 (2006)
  [hep-th/0505041].

\bibitem{Weigand:2006yj} 
  T.~Weigand,
  ``Compactifications of the heterotic string with unitary bundles,''
  Fortsch.\ Phys.\  {\bf 54}, 963 (2006).

\bibitem{Distler:2007av} 
  J.~Distler, E.~Sharpe,
  ``Heterotic compactifications with principal bundles for general groups and general levels,''
  Adv.\ Theor.\ Math.\ Phys.\  {\bf 14}, 335 (2010)
  [hep-th/0701244 [HEP-TH]].


\bibitem{cicy}
P.~Candelas, A.~M.~Dale, C.~A.~Lutken, R.~Schimmrigk,
  ``Complete Intersection Calabi-Yau Manifolds,''
  Nucl.\ Phys.\ B {\bf 298}, 493 (1988).\\
P.~Candelas, C.~A.~Lutken, R.~Schimmrigk,
``Complete Intersection Calabi-Yau Manifolds. 2. Three Generation
  Manifolds,'' 
  Nucl.\ Phys.\ B {\bf 306}, 113 (1988).\\
M.~Gagnon, Q.~Ho-Kim,
  ``An Exhaustive list of complete intersection Calabi-Yau manifolds,''
  Mod.\ Phys.\ Lett.\ A {\bf 9} (1994) 2235.  


\bibitem{Morrison:1996na} 
  D.~R.~Morrison, C.~Vafa,
  ``Compactifications of F theory on Calabi-Yau threefolds. 1 \& 2,''
  Nucl.\ Phys.\ B {\bf 473}, 74 (1996)
  [hep-th/9602114];
  Nucl.\ Phys.\ B {\bf 476}, 437 (1996)
  [hep-th/9603161].

\bibitem{Grassi:2000we} 
  A.~Grassi, D.~R.~Morrison,
  ``Group representations and the Euler characteristic of elliptically fibered Calabi-Yau threefolds,''
  math/0005196 [math-ag].


\bibitem{Donagi:1997dp} 
  R.~Y.~Donagi,
  ``Principal bundles on elliptic fibrations,''
  Asian J.\ Math {\bf 1}, 214 (1997)
  [alg-geom/9702002].



\bibitem{Gabella:2008id} 
  M.~Gabella, Y.~-H.~He, A.~Lukas,
  ``An Abundance of Heterotic Vacua,''
  JHEP {\bf 0812}, 027 (2008)
  [arXiv:0808.2142 [hep-th]].


\bibitem{BB}
Victor V.~Batyrev, Lev A.~Borisov
``On Calabi-Yau Complete Intersections in Toric Varieties'',
 arXiv:alg-geom/9412017

\bibitem{Avram:1997rs} 
  A.~C.~Avram, M.~Kreuzer, M.~Mandelberg, H.~Skarke,
  ``The Web of Calabi-Yau hypersurfaces in toric varieties,''
  Nucl.\ Phys.\ B {\bf 505}, 625 (1997)
  [hep-th/9703003].

\bibitem{Kreuzer:2000qv} 
  M.~Kreuzer, H.~Skarke,
  ``Reflexive polyhedra, weights and toric Calabi-Yau fibrations,''
  Rev.\ Math.\ Phys.\  {\bf 14}, 343 (2002)
  [math/0001106 [math-ag]].

\bibitem{Candelas:1989hd} 
  P.~Candelas, M.~Lynker, R.~Schimmrigk,
  ``Calabi-Yau Manifolds in Weighted P(4),''
  Nucl.\ Phys.\ B {\bf 341}, 383 (1990).

\bibitem{Cicoli:2011it} 
  M.~Cicoli, M.~Kreuzer, C.~Mayrhofer,
  ``Toric K3-Fibred Calabi-Yau Manifolds with del Pezzo Divisors for String Compactifications,''
  JHEP {\bf 1202}, 002 (2012)
  [arXiv:1107.0383 [hep-th]].

\bibitem{Taylor:2012dr} 
  W.~Taylor,
  ``On the Hodge structure of elliptically fibered Calabi-Yau threefolds,''
  JHEP {\bf 1208}, 032 (2012)
  [arXiv:1205.0952 [hep-th]].


\bibitem{Braun:2011ux} 
  V.~Braun,
  ``Toric Elliptic Fibrations and F-Theory Compactifications,''
  JHEP {\bf 1301}, 016 (2013)
  [arXiv:1110.4883 [hep-th]].


\bibitem{Candelas:2012uu} 
  P.~Candelas, A.~Constantin, H.~Skarke,
  ``An Abundance of K3 Fibrations from Polyhedra with Interchangeable Parts,''
  arXiv:1207.4792 [hep-th].


\bibitem{Candelas:2008wb} 
  P.~Candelas, R.~Davies,
  ``New Calabi-Yau Manifolds with Small Hodge Numbers,''
  Fortsch.\ Phys.\  {\bf 58}, 383 (2010)
  [arXiv:0809.4681 [hep-th]].

\bibitem{Braun:2009qy} 
  V.~Braun, P.~Candelas, R.~Davies,
  ``A Three-Generation Calabi-Yau Manifold with Small Hodge Numbers,''
  Fortsch.\ Phys.\  {\bf 58}, 467 (2010)
  [arXiv:0910.5464 [hep-th]].


\bibitem{Qureshi:2011cg} 
  M.~I.~Qureshi, B.~Szendroi,
  ``Calabi-Yau threefolds in weighted flag varieties,''
  Adv.\ High Energy Phys.\  {\bf 2012}, 547317 (2012)
  [arXiv:1105.4282 [math.AG]].

\bibitem{He:2010uj}
  Y.~-H.~.He,
  ``An Algorithmic Approach to Heterotic String Phenomenology,''
  Mod.\ Phys.\ Lett.\ A {\bf 25} (2010) 79
  [arXiv:1001.2419 [hep-th]].


\bibitem{Horava:1996ma}
  P.~Horava, E.~Witten,
  ``Eleven-dimensional supergravity on a manifold with boundary,''
  Nucl.\ Phys.\ B {\bf 475}, 94 (1996)
  [hep-th/9603142].


\bibitem{Braun:2005nv} 
  V.~Braun, Y.~-H.~He, B.~A.~Ovrut, T.~Pantev,
  ``The Exact MSSM spectrum from string theory,''
  JHEP {\bf 0605}, 043 (2006)
  [hep-th/0512177].

\bibitem{Bouchard:2005ag} 
  V.~Bouchard, R.~Donagi,
  ``An SU(5) heterotic standard model,''
  Phys.\ Lett.\ B {\bf 633}, 783 (2006)
  [hep-th/0512149].


\bibitem{monad}
G.~Horrocks, ``Vector bundles on the punctured spectrum of a local ring'', 
Proc.~LMS, 1964, 14 (4): 689–713\\
W.~Barth, K.~Hulek, ``Monads and moduli of vector bundles'', 
Manuscripta Mathematica 25 (4): 323–347, 1978.

\bibitem{Anderson:2007nc} 
  L.~B.~Anderson, Y.~-H.~He, A.~Lukas,
  ``Heterotic Compactification, An Algorithmic Approach,''
  JHEP {\bf 0707}, 049 (2007)
  [hep-th/0702210 [HEP-TH]].

\bibitem{Anderson:2008uw} 
  L.~B.~Anderson, Y.~-H.~He, A.~Lukas,
  ``Monad Bundles in Heterotic String Compactifications,''
  JHEP {\bf 0807}, 104 (2008)
  [arXiv:0805.2875 [hep-th]].

\bibitem{Anderson:2009ge} 
  L.~B.~Anderson, J.~Gray, D.~Grayson, Y.~-H.~He, A.~Lukas,
  ``Yukawa Couplings in Heterotic Compactification,''
  Commun.\ Math.\ Phys.\  {\bf 297}, 95 (2010)
  [arXiv:0904.2186 [hep-th]].


\bibitem{Anderson:2009mh} 
  L.~B.~Anderson, J.~Gray, Y.~-H.~He, A.~Lukas,
  ``Exploring Positive Monad Bundles And A New Heterotic Standard Model,''
  JHEP {\bf 1002}, 054 (2010)
  [arXiv:0911.1569 [hep-th]].

\bibitem{Anderson:2012yf} 
  L.~B.~Anderson, J.~Gray, A.~Lukas, E.~Palti,
  ``Heterotic Line Bundle Standard Models,''
  JHEP {\bf 1206}, 113 (2012)
  [arXiv:1202.1757 [hep-th]].

\bibitem{Anderson:2011ns} 
  L.~B.~Anderson, J.~Gray, A.~Lukas, E.~Palti,
  ``Two Hundred Heterotic Standard Models on Smooth Calabi-Yau Threefolds,''
  Phys.\ Rev.\ D {\bf 84}, 106005 (2011)
  [arXiv:1106.4804 [hep-th]].


\bibitem{Anderson:2013xx} 
  L.~B.~Anderson, A.~Constantin, J.~Gray, A.~Lukas and E.~Palti,
  ``A Comprehensive Scan for Heterotic SU(5) GUT models,''
  arXiv:1307.4787 [hep-th].

\bibitem{Anderson:2009nt} 
  L.~B.~Anderson, J.~Gray, A.~Lukas, B.~Ovrut,
  ``Stability Walls in Heterotic Theories,''
  JHEP {\bf 0909}, 026 (2009)
  [arXiv:0905.1748 [hep-th]].

\bibitem{He:2012iy} 
  Y.~-H.~He, S.~-J.~Lee,
  ``Quiver Structure of Heterotic Moduli,''
  JHEP {\bf 1211}, 119 (2012)
  [arXiv:1208.3004 [hep-th]].


\bibitem{He:2009wi} 
  Y.~-H.~He, S.~-J.~Lee, A.~Lukas,
  ``Heterotic Models from Vector Bundles on Toric Calabi-Yau Manifolds,''
  JHEP {\bf 1005}, 071 (2010)
  [arXiv:0911.0865 [hep-th]].


\bibitem{He:2011rs} 
  Y.~-H.~He, M.~Kreuzer, S.~-J.~Lee, A.~Lukas,
  ``Heterotic Bundles on Calabi-Yau Manifolds with Small Picard Number,''
  JHEP {\bf 1112}, 039 (2011)
  [arXiv:1108.1031 [hep-th]].

\bibitem{Gray:2012jy} 
  J.~Gray, Y.~-H.~He, V.~Jejjala, B.~Jurke, B.~D.~Nelson, J.~Simon,
  ``Calabi-Yau Manifolds with Large Volume Vacua,''
  Phys.\ Rev.\ D {\bf 86}, 101901 (2012)
  [arXiv:1207.5801 [hep-th]].

\bibitem{Gao:2013pra} 
  X.~Gao, P.~Shukla,
  ``On Classifying the Divisor Involutions in Calabi-Yau Threefolds,''
  arXiv:1307.1139 [hep-th].

\bibitem{Blumenhagen:2011xn} 
  R.~Blumenhagen, B.~Jurke, T.~Rahn,
  ``Computational Tools for Cohomology of Toric Varieties,''
  Adv.\ High Energy Phys.\  {\bf 2011}, 152749 (2011)
  [arXiv:1104.1187 [hep-th]].

\bibitem{Maldacena:1997re} 
  J.~M.~Maldacena,
  ``The Large N limit of superconformal field theories and supergravity,''
  Adv.\ Theor.\ Math.\ Phys.\  {\bf 2}, 231 (1998)
  [hep-th/9711200].


\bibitem{Uranga:2007zza} 
  A.~M.~Uranga,
  ``The standard model in string theory from D-branes,''
  Nucl.\ Phys.\ Proc.\ Suppl.\  {\bf 171}, 119 (2007).

\bibitem{Verlinde:2005jr} 
  H.~Verlinde, M.~Wijnholt,
  ``Building the standard model on a D3-brane,''
  JHEP {\bf 0701}, 106 (2007)
  [hep-th/0508089].

\bibitem{Gmeiner:2005vz} 
  F.~Gmeiner, R.~Blumenhagen, G.~Honecker, D.~Lust, T.~Weigand,
  ``One in a billion: MSSM-like D-brane statistics,''
  JHEP {\bf 0601}, 004 (2006)
  [hep-th/0510170].

\bibitem{Antoniadis:2002qm} 
  I.~Antoniadis, E.~Kiritsis, J.~Rizos, T.~N.~Tomaras,
  ``D-branes and the standard model,''
  Nucl.\ Phys.\ B {\bf 660}, 81 (2003)
  [hep-th/0210263].

\bibitem{Berenstein:2001nk} 
  D.~Berenstein, V.~Jejjala, R.~G.~Leigh,
  ``The Standard model on a D-brane,''
  Phys.\ Rev.\ Lett.\  {\bf 88}, 071602 (2002)
  [hep-ph/0105042].

\bibitem{Bailin:2000kd} 
  D.~Bailin, G.~V.~Kraniotis, A.~Love,
  ``Supersymmetric standard models on D-branes,''
  Phys.\ Lett.\ B {\bf 502}, 209 (2001)
  [hep-th/0011289].

\bibitem{Aldazabal:2000sa} 
  G.~Aldazabal, L.~E.~Ibanez, F.~Quevedo, A.~M.~Uranga,
  ``D-branes at singularities: A Bottom up approach to the string embedding of the standard model,''
  JHEP {\bf 0008}, 002 (2000)
  [hep-th/0005067].

\bibitem{Katz:1996fh} 
  S.~H.~Katz, A.~Klemm, C.~Vafa,
  ``Geometric engineering of quantum field theories,''
  Nucl.\ Phys.\ B {\bf 497}, 173 (1997)
  [hep-th/9609239].


\bibitem{Lawrence:1998ja} 
  A.~E.~Lawrence, N.~Nekrasov, C.~Vafa,
  ``On conformal field theories in four-dimensions,''
  Nucl.\ Phys.\ B {\bf 533}, 199 (1998)
  [hep-th/9803015].

\bibitem{Kachru:1998ys}
  S.~Kachru, E.~Silverstein,
  ``4-D conformal theories and strings on orbifolds,''
  Phys.\ Rev.\ Lett.\  {\bf 80} (1998) 4855
  [hep-th/9802183].

\bibitem{Johnson:1996py} 
  C.~V.~Johnson, R.~C.~Myers,
  ``Aspects of type IIB theory on ALE spaces,''
  Phys.\ Rev.\ D {\bf 55}, 6382 (1997)
  [hep-th/9610140].

\bibitem{He:2004rn} 
  Y.~-H.~He,
  ``Lectures on D-branes, gauge theories and Calabi-Yau singularities,''
  hep-th/0408142.

\bibitem{duVal}
P.~du Val, 
``On isolated singularities of surfaces which do not affect the conditions of adjunction. I,II,III'', 
Proc.~Cam.~Phil.~Soc.~30 (4) 1934.

\bibitem{mckay} 
J.~McKay, ``Graphs, Singularities, and Finite Groups,'' Proc. Symp. Pure Math.
Vol 37, 183-186 (1980).

\bibitem{blich}
H.~F.~Blichfeldt, ``Finite Collineation Groups,'' Univ. Chicago Press,
Chicago, 1917.

\bibitem{Hanany:1998sd} 
  A.~Hanany, Y.~-H.~He,
  ``NonAbelian finite gauge theories,''
  JHEP {\bf 9902}, 013 (1999)
  [hep-th/9811183].

\bibitem{Hanany:1999sp} 
  A.~Hanany, Y.~-H.~He,
  JHEP {\bf 0102}, 027 (2001)
  [hep-th/9905212].


\bibitem{duca}
Pasquale Del Pezzo, Duca di Cajanello, {\it Opera Omnia}.
cf.~ G. Gallucci, {\it Rend. R. Acc. delle Scienze Fisiche e Mat. di
Napoli}, 8, 1938, 162-167. 


\bibitem{Iqbal:2001ye} 
  A.~Iqbal, A.~Neitzke, C.~Vafa,
  ``A Mysterious duality,''
  Adv.\ Theor.\ Math.\ Phys.\  {\bf 5}, 769 (2002)
  [hep-th/0111068].

\bibitem{Herzog:2003dj} 
  C.~P.~Herzog, J.~Walcher,
  ``Dibaryons from exceptional collections,''
  JHEP {\bf 0309}, 060 (2003)
  [hep-th/0306298].

\bibitem{Herzog:2003zc} 
  C.~P.~Herzog,
  ``Exceptional collections and del Pezzo gauge theories,''
  JHEP {\bf 0404}, 069 (2004)
  [hep-th/0310262].

\bibitem{Herzog:2005sy} 
  C.~P.~Herzog, R.~L.~Karp,
  ``Exceptional collections and D-branes probing toric singularities,''
  JHEP {\bf 0602}, 061 (2006)
  [hep-th/0507175].

\bibitem{Hanany:2006nm} 
  A.~Hanany, C.~P.~Herzog, D.~Vegh,
  ``Brane tilings and exceptional collections,''
  JHEP {\bf 0607}, 001 (2006)
  [hep-th/0602041].


\bibitem{Aspinwall:2004vm} 
  P.~S.~Aspinwall, I.~V.~Melnikov,
  ``D-branes on vanishing del Pezzo surfaces,''
  JHEP {\bf 0412}, 042 (2004)
  [hep-th/0405134].


\bibitem{Wijnholt:2002qz} 
  M.~Wijnholt,
  ``Large volume perspective on branes at singularities,''
  Adv.\ Theor.\ Math.\ Phys.\  {\bf 7}, 1117 (2004)
  [hep-th/0212021].


\bibitem{Dolan:2011qu} 
  M.~J.~Dolan, S.~Krippendorf, F.~Quevedo,
  ``Towards a Systematic Construction of Realistic D-brane Models on a del Pezzo Singularity,''
  JHEP {\bf 1110}, 024 (2011)
  [arXiv:1106.6039 [hep-th]].

\bibitem{Cicoli:2012vw} 
  M.~Cicoli, S.~Krippendorf, C.~Mayrhofer, F.~Quevedo, R.~Valandro,
  ``D-Branes at del Pezzo Singularities: Global Embedding and Moduli Stabilisation,''
  JHEP {\bf 1209}, 019 (2012)
  [arXiv:1206.5237 [hep-th]].


\bibitem{Malyshev:2007yb} 
  D.~Malyshev,
  ``Del Pezzo singularities and SUSY breaking,''
  Adv.\ High Energy Phys.\  {\bf 2011}, 630892 (2011)
  [arXiv:0705.3281 [hep-th]].

\bibitem{KN}
B.~Karpov, D.~Nogin,
``Three-block exceptional collections over Del Pezzo surfaces'',
 arXiv:alg-geom/9703027.

\bibitem{Benvenuti:2004dw} 
  S.~Benvenuti, A.~Hanany,
  ``New results on superconformal quivers,''
  JHEP {\bf 0604}, 032 (2006)
  [hep-th/0411262].

\bibitem{Hanany:2012mb} 
  A.~Hanany, Y.~-H.~He, C.~Sun, S.~Sypsas,
  ``Superconformal Block Quivers, Duality Trees and Diophantine Equations,''
  arXiv:1211.6111 [hep-th].



\bibitem{Bouchard:2007ik} 
  V.~Bouchard,
  ``Lectures on complex geometry, Calabi-Yau manifolds and toric geometry,''
  hep-th/0702063 [HEP-TH].


\bibitem{Witten:1993yc} 
  E.~Witten,
  ``Phases of N=2 theories in two-dimensions,''
  Nucl.\ Phys.\ B {\bf 403}, 159 (1993)
  [hep-th/9301042].

\bibitem{Douglas:1997de} 
  M.~R.~Douglas, B.~R.~Greene, D.~R.~Morrison,
  ``Orbifold resolution by D-branes,''
  Nucl.\ Phys.\ B {\bf 506}, 84 (1997)
  [hep-th/9704151].

\bibitem{Beasley:1999uz} 
  C.~Beasley, B.~R.~Greene, C.~I.~Lazaroiu, M.~R.~Plesser,
  ``D3-branes on partial resolutions of Abelian quotient singularities of Calabi-Yau threefolds,''
  Nucl.\ Phys.\ B {\bf 566}, 599 (2000)
  [hep-th/9907186].

\bibitem{Feng:2000mi} 
  B.~Feng, A.~Hanany, Y.~-H.~He,
  ``D-brane gauge theories from toric singularities and toric duality,''
  Nucl.\ Phys.\ B {\bf 595}, 165 (2001)
  [hep-th/0003085].

\bibitem{Feng:2001xr} 
  B.~Feng, A.~Hanany, Y.~-H.~He,
  ``Phase structure of D-brane gauge theories and toric duality,''
  JHEP {\bf 0108}, 040 (2001)
  [hep-th/0104259].


\bibitem{Feng:2001bn} 
  B.~Feng, A.~Hanany, Y.~-H.~He, A.~M.~Uranga,
  ``Toric duality as Seiberg duality and brane diamonds,''
  JHEP {\bf 0112}, 035 (2001)
  [hep-th/0109063].

\bibitem{Gauntlett:2004hh} 
  J.~P.~Gauntlett, D.~Martelli, J.~F.~Sparks, D.~Waldram,
  ``A New infinite class of Sasaki-Einstein manifolds,''
  Adv.\ Theor.\ Math.\ Phys.\  {\bf 8}, 987 (2006)
  [hep-th/0403038].

\bibitem{Benvenuti:2004dy} 
  S.~Benvenuti, S.~Franco, A.~Hanany, D.~Martelli, J.~Sparks,
  ``An Infinite family of superconformal quiver gauge theories with Sasaki-Einstein duals,''
  JHEP {\bf 0506}, 064 (2005)
  [hep-th/0411264].

\bibitem{Beasley:2001zp} 
  C.~E.~Beasley, M.~R.~Plesser,
  ``Toric duality is Seiberg duality,''
  JHEP {\bf 0112}, 001 (2001)
  [hep-th/0109053].

\bibitem{binomial}
D.~Eisenbud, B.~Sturmfels, 
``Binomial Ideals'',
[alg-geom/9401001].

\bibitem{Cachazo:2001gh}
  F.~Cachazo, S.~Katz, C.~Vafa,
  ``Geometric transitions and N=1 quiver theories,''
  hep-th/0108120.

\bibitem{Cachazo:2001sg} 
  F.~Cachazo, B.~Fiol, K.~A.~Intriligator, S.~Katz, C.~Vafa,
  ``A Geometric unification of dualities,''
  Nucl.\ Phys.\ B {\bf 628}, 3 (2002)
  [hep-th/0110028].

\bibitem{Franco:2004jz} 
  S.~Franco, Y.~-H.~He, C.~Herzog, J.~Walcher,
  ``Chaotic duality in string theory,''
  Phys.\ Rev.\ D {\bf 70}, 046006 (2004)
  [hep-th/0402120].

\bibitem{FZ}
S.~Fomin, A.~Zelevinsky, 
``Cluster algebras. I. Foundations.'' 
J. Amer. Math. Soc., 15 (2), 2002.

\bibitem{ArkaniHamed:2012nw} 
  N.~Arkani-Hamed, J.~L.~Bourjaily, F.~Cachazo, A.~B.~Goncharov, A.~Postnikov, J.~Trnka,
  ``Scattering Amplitudes and the Positive Grassmannian,''
  arXiv:1212.5605 [hep-th].


\bibitem{Hanany:2005ve} 
  A.~Hanany, K.~D.~Kennaway,
  ``Dimer models and toric diagrams,''
  hep-th/0503149.

\bibitem{Franco:2005rj} 
  S.~Franco, A.~Hanany, K.~D.~Kennaway, D.~Vegh, B.~Wecht,
  ``Brane dimers and quiver gauge theories,''
  JHEP {\bf 0601}, 096 (2006)
  [hep-th/0504110].

\bibitem{Franco:2005sm} 
  S.~Franco, A.~Hanany, D.~Martelli, J.~Sparks, D.~Vegh, B.~Wecht,
  ``Gauge theories from toric geometry and brane tilings,''
  JHEP {\bf 0601}, 128 (2006)
  [hep-th/0505211].

\bibitem{Feng:2005gw} 
  B.~Feng, Y.~-H.~He, K.~D.~Kennaway, C.~Vafa,
  ``Dimer models from mirror symmetry and quivering amoebae,''
  Adv.\ Theor.\ Math.\ Phys.\  {\bf 12}, 489 (2008)
  [hep-th/0511287].

\bibitem{GarciaEtxebarria:2006aq} 
  I.~Garcia-Etxebarria, F.~Saad, A.~M.~Uranga,
  ``Quiver gauge theories at resolved and deformed singularities using dimers,''
  JHEP {\bf 0606}, 055 (2006)
  [hep-th/0603108].

\bibitem{Kennaway:2007tq} 
  K.~D.~Kennaway,
  ``Brane Tilings,''
  Int.\ J.\ Mod.\ Phys.\ A {\bf 22}, 2977 (2007)
  [arXiv:0706.1660 [hep-th]].

\bibitem{Yamazaki:2008bt} 
  M.~Yamazaki,
  ``Brane Tilings and Their Applications,''
  Fortsch.\ Phys.\  {\bf 56}, 555 (2008)
  [arXiv:0803.4474 [hep-th]].

\bibitem{He:2012js} 
  Y.~-H.~He,
  ``Bipartita: Physics, Geometry \& Number Theory,''
  arXiv:1210.4388 [hep-th].


\bibitem{He:2012kw} 
  Y.~-H.~He, J.~McKay,
  ``N=2 Gauge Theories: Congruence Subgroups, Coset Graphs and Modular Surfaces,''
  J.\ Math.\ Phys.\  {\bf 54}, 012301 (2013)
  [J.\ Math.\ Phys.\  {\bf 54}, 012301 (2013)]
  [arXiv:1201.3633 [hep-th]].

\bibitem{Hanany:2011ra} 
  A.~Hanany, Y.~-H.~He, V.~Jejjala, J.~Pasukonis, S.~Ramgoolam, D.~Rodriguez-Gomez,
  ``The Beta Ansatz: A Tale of Two Complex Structures,''
  JHEP {\bf 1106}, 056 (2011)
  [arXiv:1104.5490 [hep-th]].


\bibitem{Jejjala:2010vb} 
  V.~Jejjala, S.~Ramgoolam, D.~Rodriguez-Gomez,
  ``Toric CFTs, Permutation Triples and Belyi Pairs,''
  JHEP {\bf 1103}, 065 (2011)
  [arXiv:1012.2351 [hep-th]].

\bibitem{Ashok:2006br} 
  S.~K.~Ashok, F.~Cachazo, E.~Dell'Aquila,
  ``Children's drawings from Seiberg-Witten curves,''
  Commun.\ Num.\ Theor.\ Phys.\  {\bf 1}, 237 (2007)
  [hep-th/0611082].


\bibitem{Witten:2000nv} 
  E.~Witten,
  ``Supersymmetric index in four-dimensional gauge theories,''
  Adv.\ Theor.\ Math.\ Phys.\  {\bf 5}, 841 (2002)
  [hep-th/0006010].
  
\bibitem{Romelsberger:2005eg} 
  C.~Romelsberger,
  ``Counting chiral primaries in N = 1, d=4 superconformal field theories,''
  Nucl.\ Phys.\ B {\bf 747}, 329 (2006)
  [hep-th/0510060].
  
\bibitem{Bhattacharya:2008zy} 
  J.~Bhattacharya, S.~Bhattacharyya, S.~Minwalla, S.~Raju,
  ``Indices for Superconformal Field Theories in 3,5 and 6 Dimensions,''
  JHEP {\bf 0802}, 064 (2008)
  [arXiv:0801.1435 [hep-th]].
  
\bibitem{Gadde:2011uv} 
  A.~Gadde, L.~Rastelli, S.~S.~Razamat, W.~Yan,
  ``Gauge Theories and Macdonald Polynomials,''
  arXiv:1110.3740 [hep-th].

\bibitem{Benvenuti:2006qr} 
  S.~Benvenuti, B.~Feng, A.~Hanany, Y.~-H.~He,
  ``Counting BPS Operators in Gauge Theories: Quivers, Syzygies and Plethystics,''
  JHEP {\bf 0711}, 050 (2007)
  [hep-th/0608050].

\bibitem{Feng:2007ur} 
  B.~Feng, A.~Hanany, Y.~-H.~He,
  ``Counting gauge invariants: The Plethystic program,''
  JHEP {\bf 0703}, 090 (2007)
  [hep-th/0701063].

\bibitem{stanley}
R.~Stanley, 
``Hilbert functions of graded algebras,'' 
Adv. Math. 28 (1978), 57-83.

\bibitem{Forcella:2008bb} 
  D.~Forcella, A.~Hanany, Y.~-H.~He, A.~Zaffaroni,
  ``The Master Space of N=1 Gauge Theories,''
  JHEP {\bf 0808}, 012 (2008)
  [arXiv:0801.1585 [hep-th]].




\bibitem{Mehta:2012wk} 
  D.~Mehta, Y.~-H.~He, J.~D.~Hauenstein,
  ``Numerical Algebraic Geometry: A New Perspective on String and Gauge Theories,''
  JHEP {\bf 1207}, 018 (2012)
  [arXiv:1203.4235 [hep-th]].


\bibitem{Hauenstein:2012xs} 
  J.~Hauenstein, Y.~-H.~He, D.~Mehta,
  ``Numerical Analyses on Moduli Space of Vacua,''
  arXiv:1210.6038 [hep-th].

\bibitem{He:2013yk} 
  Y.~-H.~He, D.~Mehta, M.~Niemerg, M.~Rummel, A.~Valeanu,
  ``Exploring the Potential Energy Landscape Over a Large Parameter-Space,''
  arXiv:1301.0946 [hep-th].




\bibitem{Gray:2005sr} 
  J.~Gray, Y.~-H.~He, V.~Jejjala, B.~D.~Nelson,
  ``Vacuum geometry and the search for new physics,''
  Phys.\ Lett.\ B {\bf 638}, 253 (2006)
  [hep-th/0511062].

\bibitem{Gray:2006jb} 
  J.~Gray, Y.~-H.~He, V.~Jejjala, B.~D.~Nelson,
  ``Exploring the vacuum geometry of N=1 gauge theories,''
  Nucl.\ Phys.\ B {\bf 750}, 1 (2006)
  [hep-th/0604208].


\bibitem{Gray:2008zs} 
  J.~Gray, Y.~-H.~He, A.~Ilderton, A.~Lukas,
  ``STRINGVACUA: A Mathematica Package for Studying Vacuum Configurations in String Phenomenology,''
  Comput.\ Phys.\ Commun.\  {\bf 180}, 107 (2009)
  [arXiv:0801.1508 [hep-th]].

\bibitem{Gray:2007yq} 
  J.~Gray, Y.~-H.~He, A.~Ilderton, A.~Lukas,
  ``A New Method for Finding Vacua in String Phenomenology,''
  JHEP {\bf 0707}, 023 (2007)
  [hep-th/0703249 [HEP-TH]].

\bibitem{Gray:2009fy} 
  J.~Gray,
  ``A Simple Introduction to Grobner Basis Methods in String Phenomenology,''
  Adv.\ High Energy Phys.\  {\bf 2011}, 217035 (2011)
  [arXiv:0901.1662 [hep-th]].



\bibitem{Gherghetta:1995dv} 
  T.~Gherghetta, C.~F.~Kolda, S.~P.~Martin,
  ``Flat directions in the scalar potential of the supersymmetric standard model,''
  Nucl.\ Phys.\ B {\bf 468}, 37 (1996)
  [hep-ph/9510370].


\bibitem{Gray:2008yu} 
  J.~Gray, A.~Hanany, Y.~-H.~He, V.~Jejjala, N.~Mekareeya,
  ``SQCD: A Geometric Apercu,''
  JHEP {\bf 0805}, 099 (2008)
  [arXiv:0803.4257 [hep-th]].


\bibitem{Hanany:2010vu} 
  A.~Hanany, E.~E.~Jenkins, A.~V.~Manohar, G.~Torri,
  ``Hilbert Series for Flavor Invariants of the Standard Model,''
  JHEP {\bf 1103}, 096 (2011)
  [arXiv:1010.3161 [hep-ph]].


\bibitem{Hanany:2008sb} 
  A.~Hanany, N.~Mekareeya, G.~Torri,
  ``The Hilbert Series of Adjoint SQCD,''
  Nucl.\ Phys.\ B {\bf 825}, 52 (2010)
  [arXiv:0812.2315 [hep-th]].

\bibitem{Basor:2011da} 
  E.~Basor, Y.~Chen, N.~Mekareeya,
  ``The Hilbert Series of N=1 $SO(N_c)$ and $Sp(N_c)$ SQCD, Painlev\'e VI and Integrable Systems,''
  Nucl.\ Phys.\ B {\bf 860}, 421 (2012)
  [arXiv:1112.3848 [hep-th]].
}
\end{thebibliography}
\end{document}